\documentclass[onecolumn,12pt]{IEEEtran}
\usepackage{mathtools}
\usepackage{amssymb}
\usepackage{amsmath,color,amsthm}
\usepackage{breqn}
\usepackage{bm}
\usepackage{cite}
\usepackage{enumitem}
\usepackage{caption}
\usepackage{lipsum}
\usepackage{hyperref}
\usepackage{cuted}
\usepackage{enumerate}
\usepackage{booktabs} 
\usepackage{graphicx}
\usepackage{float}
\usepackage{mathrsfs}

\usepackage{algpseudocode}
\usepackage{algorithm}
\algnewcommand\algorithmicforeach{\textbf{for each}}
\algdef{S}[FOR]{ForEach}[1]{\algorithmicforeach\ #1\ \algorithmicdo}
\usepackage{eqparbox}
\newdimen{\algindent}
\algnewcommand\LeftComment[2]{%
\hspace{#1\algindent}$\triangleright$ \eqparbox{COMMENT}{#2} \hfill %
}
\algnewcommand\LeftCommentNoTriangle[2]{%
\hspace{#1\algindent} \eqparbox{COMMENT}{#2} \hfill %
}

\usepackage{amsmath,amsthm,amssymb}
\DeclareMathOperator*{\argmax}{argmax}

\newtheorem{axiom}{Axiom}

\usepackage{wasysym} 

\DeclarePairedDelimiterX\Basics[1](){ #1}

\algblock{ParFor}{EndParFor}
\algnewcommand\algorithmicparfor{\textbf{parfor}}
\algnewcommand\algorithmicpardo{\textbf{do}}
\algnewcommand\algorithmicendparfor{\textbf{end\ parfor}}
\algrenewtext{ParFor}[1]{\algorithmicparfor\ #1\ \algorithmicpardo}
\algrenewtext{EndParFor}{\algorithmicendparfor}

\begin{document}
	\title{Self-Organizing mmWave MIMO Cell-Free Networks With Hybrid Beamforming: A Hierarchical DRL-Based Design} 
\author{Yasser Al-Eryani, {\it Student Member, IEEE} and Ekram Hossain, {\it Fellow, IEEE} \thanks{Y. Al-Eryani and E. Hossain are with the Department of Electrical and Computer Engineering at the University of Manitoba, Canada (emails: aleryany@myumanitoba.ca, Ekram.Hossain@umanitoba.ca). The work was supported by a Discovery Grant form the Natural Sciences and Engineering Research Council of Canada (NSERC).}}
 	\maketitle
	\begin{abstract}
In a cell-free wireless network, distributed access points (APs)  jointly serve all user equipments (UEs) within the their coverage area by using the same time/frequency resources. 
In this paper, we develop a novel downlink cell-free multiple-input multiple-output (MIMO) millimeter wave (mmWave) network architecture that enables all APs and UEs to dynamically self-partition into a set of independent cell-free subnetworks in a time-slot basis. For this, we propose several  network partitioning algorithms based on deep reinforcement learning (DRL). Furthermore, to mitigate interference between different cell-free subnetworks, we develop a novel hybrid analog beamsteering-digital beamforming model that zero-forces interference among cell-free subnetworks and at the same time maximizes the instantaneous sum-rate of all UEs within each subnetwork. Specifically, the hybrid beamforming model is implemented by using a novel mixed DRL-convex optimization method in which analog beamsteering between APs and UEs is conducted based on DRL  while digital beamforming is modeled and solved as a convex optimization problem. The DRL models for network clustering and hybrid beamsteering are combined into a single hierarchical DRL design that enables exchange of DRL agents' experiences during both network training and operation. We also benchmark the performance of DRL models for clustering and beamsteering  in terms of network performance, convergence rate, and computational complexity. Results show a significant rate enhancement and complexity reduction of the proposed hybrid beamforming scheme compared to its conventional all-digital counterpart. This performance enhancement becomes more significant as the number of network partitions increases. For DRL-based network clustering, the policy gradient (PG) algorithm offers the best possible performance in terms of stability and convergence rate while the state-action-reward-state-action (SARSA) algorithm suffers from significant variance, slower convergence, and slightly inferior performance than other algorithms. For DRL-based  beamsteering, the soft actor-critic (SAC) algorithm with continuous action space shows the best performance. Also,  online training of the agents  with varying channel state information (CSI) is observed to increase the variance of the Q-values and decrease the convergence rate, with no significant effect on the average reward.
\end{abstract}
\section{Introduction}

\subsection{Background and Related Work}
Cell-free (or cell-less) network architectures are envisioned to  increased coverage and transmission rates in future generation wireless systems~\cite{Cell_Less_2}. In a cell-free  wireless network, a large number of user equipments (UEs) in a geographical area will be served simultaneously by a large number of distributed access points (APs) based on non-orthogonal multiple access.  The distributed APs in a cell-free system coordinate/cooperate with each other through a centralized processing pool \cite{Dynamic_Cell_Free} for estimating channel state information (CSI) \cite{DRL_MEC_1,Estimation_2}, uplink (downlink) decoding (beamforming) \cite{Beamforming_Algorithms_1,Beamforming_2,Cell_Less_Beamforming_2}, and improving transmission performance \cite{GCoMP,Cell_Free_Complexity_1,Dynamic_Cell_Free}. 
The majority of the works  on cell-free systems in the literature tackle the following major technical challenges: i) pilot contamination, ii) high computational and hardware complexity of centralized processing, and iii) traffic/signaling overhead. 
For instance, in \cite{Estimation_2}, the authors  designed a joint uplink/downlink pilot training scheme that uses orthogonal subsets of pilots in the downlink instead of using channel reciprocity concept. In \cite{estimation_3}, the authors developed a semi-blind channel estimation of uplink cell-free massive MIMO network utilizing an enhanced $K$-means clustering algorithm. 
In \cite{Beamforming_Algorithms_1}, the authors  proposed a downlink conjugate beamforming and zero-forcing (ZF) precoding scheme for a fully centralized downlink cell-free network. It was shown that the ZF technique outperforms the conjugate beamforming method at the expense of increased computational complexity. However, when the number of UEs and/or APs increases, the complexity of using ZF beamforming increases significantly due to the requirement of matrix inversion. Accordingly, in \cite{Cell_Less_Beamforming_2}, a modified conjugate beamforming technique was proposed that uses CSI coordination among the distributed APs. 
Different machine learning (ML) techniques were used for CSI estimation \cite{DRL_MEC_1} and beamforming \cite{Dynamic_Cell_Free} in cell-free networks.  For instance, the authors in \cite{DRL_MEC_1} developed a channel estimation technique for millimeter wave (mmWave)-enabled massive cell-free network using a supervised learning-based denoising convolutional neural network. The authors in \cite{Dynamic_Cell_Free}  formulated and solved a joint problem for AP clustering and uplink beamforming in a massive cell-free network using deep reinforcement learning (DRL) techniques. 
 
To reduce the complexity of centralized data processing in a cell-free network, the authors in \cite{GCoMP} proposed a user-centric partitioning method\footnote{We use the terms clustering and partitioning interchangeably.}. The proposed method also uses multi-level successive interference cancellation (SIC) at each receiver. Another low-complexity design for cell-free networks was proposed in \cite{Dynamic_Cell_Free}. The core idea is to reduce the dimensionality of beamforming matrices by dynamic clustering of APs. Each cluster then represents a single multi-antenna AP. 
All of the aforementioned low-complexity designs, however, sacrifice the performance gain of centralized processing. The complexity of solving the beamforming problem in a centralized manner (e.g. to obtain the beamforming vectors at a centralized processing unit) can however be reduced by using a distributed learning and/or processing approach while the detection of the  transmitted data is still performed at the central unit. Such a solution has been recently investigated in  \cite{Distributed_DRL_1,DRL_MEC_3}.
Specifically, in \cite{DRL_MEC_3}, the authors utilized supervised learning to solve the beamforming problem in a cell-free network by using a neural network optimizer in each AP. 
 

Along with cell-free network architectures, mmWave transmissions can be used to improve network capacity \cite{Cell_Free_mmWave_2,Cell_Free_mmWave}. Interestingly, cell-free networks were found to provide an efficient solution for the poor scattering  and high path-loss problem of mmWave transmissions \cite{Cell_Free_mmWave_2}. Due to the propagation issues related to mmWave channels, massive multiple-input multiple-output (MIMO) systems are usually used for mmWave-supported UEs \cite{mmWave_MIMO_1}. 
A mmWave and MIMO-based cell-free network, however, requires very significant amount of computational capabilities at the central unit, especially when the number of UEs within the network increases.  
\subsection{Motivation and Contributions} 


For a cell-free network, the complexity of solving the beamforming problem in a centralized manner can be reduced by partitioning the network into a group of cell-free subnetworks, each with independent set of APs and UEs. However, fixed partitioning will not be performance-efficient under fast-varying channel conditions and varying number of UEs per unit area. 
Therefore, dynamic partitioning into subnetworks based on current network and channel status will be desirable, and for practical feasibility, low-complexity solutions will be required. This motivates us to design a novel mmWave MIMO cell-free network architecture based on dynamic partitioning (or clustering) along with a hybrid analog-digital downlink beamforming method by using DRL techniques. The proposed design provides us with efficient and implementation-friendly solutions.

The main contributions of this paper can be summarized as follows:
\begin{itemize}
    \item For a mmWave MIMO cell-free network, we design a self-organizing network architecture that dynamically partitions the network into a group of subnetworks, each acting as an independent cell-free architecture.
    \item To simultaneously mitigate inter-subnetwork interference (ISNI) and intra-UE interference (IUI) while maximizing the per-UE transmission rate, we develop an innovative hybrid analog beamsteering-digital beamforming method for the proposed mmWave MIMO cell-free network.
    \item The problem of joint network partitioning, analog beamsteering, and digital beamforming is solved through a novel DRL-cum-convex optimization model. Specifically, the model consists of two interacting networks: i) one DRL model with discrete-action subspace for UE and AP clustering, ii) and another DRL model with continuous-action subspace used for analog beamsteering, the first step of the proposed hybrid beamforming method. The second step of digital beamforming is formulated and solved as a convex optimization problem within the environment of the DRL agent for analog beamsteering. 
    \item For network partitioning and beamforming, we benchmark several DRL algorithms and evaluate their performances under different system parameters. 
\end{itemize} 
\textbf{Table \ref{table:symbol}} presents the definitions of major system model parameters. 
The rest of the paper is organized as follows. \textbf{Sec. II} presents the proposed dynamic mmWave MIMO cell-free network architecture and the related modeling assumptions. Also, the general problem formulation to design the proposed system (e.g. clustering and beamforming design) is presented in this section.  \textbf{Sec. III} presents the beamforming scheme developed for the proposed system. The hierarchical DRL model for joint clustering and beamforming is presented in \textbf{Sec. IV}. In \textbf{Sec. V}, the complexity analysis of the proposed models and algorithms is presented. \textbf{Sec. VI} present the simulation results. \textbf{Sec. VII} concludes the paper. 
\begin{table}[h]
	\scriptsize \caption{{Definitions of major system model parameters}}
	\label{table:symbol}
	\begin{center}
		\begin{tabular}{|c|c||c|c|}
			\hline
			Parameter  & Definition & Parameter  & Definition    \\
			\hline
eAP & Enhanced Access Point & $\mathcal{N}$ & Number of possible subnetwork (or cluster) configurations  
\\
ECP & Edge Cloud Processor & $\mathcal{A}_{m_n}$ & Beamsteering matrix of eAP $m_n$   
\\
ISNI & Inter-Subnetwork Interference & $\mathcal{D}_{n,j}^{\text{A}}$ & Number of eAPs in the $n$-th subnetwork 
\\
IUI & Intra-UE Interference & $\mathcal{D}_{n,j}^{\text{U}}$ & Number of UEs in the $n$-th Subnetwork
\\
NCC & Network Cloud Controller & $\bm{w}_{k_n m_n}$ & Beamforming vector for  the link $m_n\rightarrow k_n$ 
\\
$\bm{B}^*$ & Hermitian transpose of a matrix $\bm{B}$ &  $m_n$ & $m$-th eAP in the $n$-th subnetwork 
\\
$\bm{B}^T$ & Transpose of a matrix $\bm{B}$ &  $k_n$ & $k$-th UE in the $n$-th subnetwork
\\
$\bm{H}_{k_n m_n}$ & CSI for the link $m_n\rightarrow k_n$ & $\bm{\mathcal{H}}_{k_n m_n}$ & Equivalent CSI for the $m_n\rightarrow k_n$ link   
\\
M & Number of eAPs & $\mathcal{S}$ & State space of a DRL model  
\\
$a$ & Number of antennas per eAP  &  $\bm{s}$ & State vector at time $t$
\\
K & Number of UEs & $\bm{s}'$ & State vector at time $t+1$ 
\\
$u$ & Number of antennas per UE & $r$ & Immediate reward of a DRL agent
\\
$N$ & Number of cell-free subnetworks  & $\bm{A}$ & Action space of DRL model  
\\
$\mathcal{L}$ & Number of mmWave paths &  $\bm{a}$ & Action vector at time $t$
\\
$\bm{\mathcal{C}}_j$ & $j$-th clustering configuration & $\bm{a}'$ & Action vector at time $t+1$  \\
\hline
		\end{tabular}
	\end{center}
\end{table}   
\section{System Model and Assumptions}
\subsection{Network Architecture}\label{Network_Architecture}
\begin{figure}[htb]
		\centering
		\includegraphics[scale = 0.2]{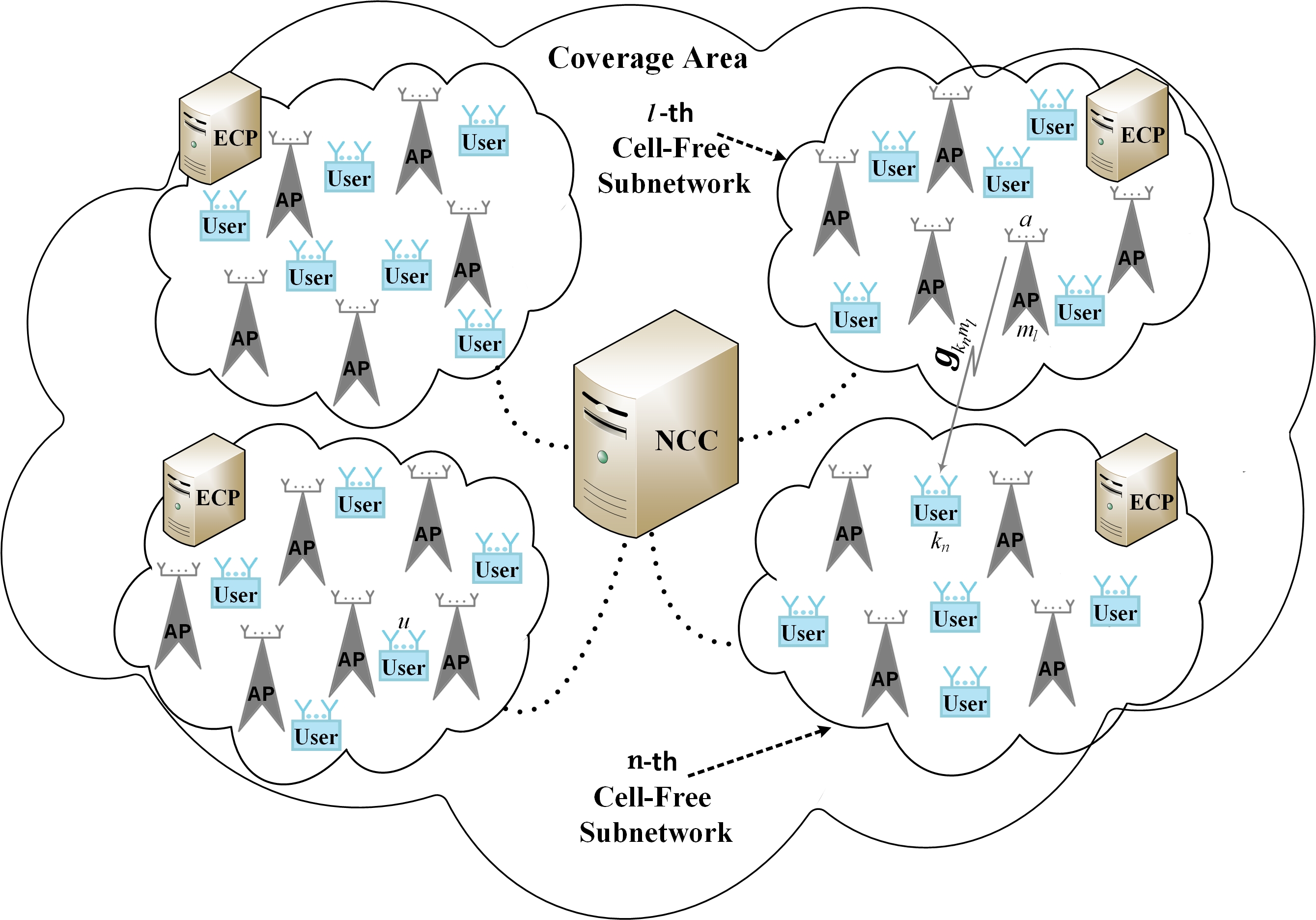}
		\caption{An example scenario for the proposed network architecture ($M = 19, K = 21$, and $N = 4$).}\label{System_Model}
	\end{figure}

We consider a downlink network with $M$ APs and $K$ UEs (\textbf{Fig.~\ref{System_Model}}). Each of the APs and UEs is assumed to be equipped with $a$ and $u$ antennas, respectively. To enable multiuser transmission, each AP is assumed to be equipped with $L$ RF chains.
All of the APs are connected to each other through fronthaul/backhaul links to form a cell-free network architecture \cite{Cell_Less_1}. 
This enables the distributed APs to collaborate to simultaneously serve all UEs within the network coverage area. Specifically, the APs collaborate through a network cloud controller (NCC).
Each of the APs is assumed to be equipped with a baseband processor that is capable of performing operations related to uplink channel training and downlink beamforming of signals transmitted to different UEs. Such an AP is referred to as an ``enhanced-AP" (eAP) to distinguish it from conventional APs with passband transmission/reception functionalities.  
In the proposed network architecture, it is assumed that the all of the eAPs and UEs are partitioned  into a set of $N$, where $1 \leq N \leq M$, non-overlapping clusters (i.e. cell-free subnetworks) on a time-slot basis. All of the UEs of a certain subnetwork are served by all of the eAPs of that subnetwork  using the same time-frequency resources. Accordingly, the number of RF chains required at each eAP will be equal to the maximum allowable number of UEs per cell-free subnetwork, i.e. $L = K-N+1$\footnote{The maximum number of UEs per subnetwork may be defined based on the hardware cost/complexity of the eAPs. This will have a direct impact on the number of RF chains required per eAP and the average amount of energy consumption \cite{RF_Chain_1}.}.  
The group of all baseband processors of eAPs within each cluster can be coordinated to form an virtual edge cloud processor (ECP) unit that is responsible for performing multiuser downlink beamforming within each subnetwork, considering signals from other clusters as Inter-subnetwork interference (ISNI) components. Each eAP may act as an ECP for its subnetwork, or all eAPs of a single subnetwork may form a virtual ECP.  
Furthermore, the clustering of the cell-free network into a group of non-overlapping cell-free subnetworks is assumed to be performed centrally at the NCC. 
These two operations of network clustering and subnetwork beamforming are performed either  at each time-slot or every several time-slots, based on current CSI and  time-varying propagation characteristics of the network. 
Note that, when $N = 1$, all eAPs and UEs of the network will belong to the same subnetwork which will form a fully centralized cell-free network. On the other hand, when $N = M$, the overall architecture will act as a conventional wireless cellular network with a reuse factor of $1$. 

Let us denote by $\bm{\mathcal{C}}=\{\{\bm{\mathcal{C}}^{\text{A}}_1, \bm{\mathcal{C}}^{\text{U}}_1\} \dots \{\bm{\mathcal{C}}^{\text{A}}_j, \bm{\mathcal{C}}^{\text{U}}_j\} \dots \{\bm{\mathcal{C}}^{\text{A}}_{\mathcal{N}},\bm{\mathcal{C}}^{\text{U}}_{\mathcal{N}} \} \}$ the set of all possible AP-UE clustering configurations such that every cluster contains at least one AP and one UE. $\mathcal{N}$ is the total number of possible clustering configurations which is a function of $M$, $K$, and $N$, i.e. $\mathcal{N} = \Theta(M,K,N)$ (to be defined in subsequent sections). 
As an example, with $M = 4$, $K = 3$, and $N = 2$, one possible set is
\[
\bm{\mathcal{C}}_j=\{\underbrace{  \{\{ \text{AP}_1, \text{AP}_3, \text{AP}_4 \}}_{\bm{\mathcal{C}}_{1,j}^{\text{A}}},\underbrace{\{\text{UE}_2\}}_{\bm{\mathcal{C}}_{1,j}^{\text{U}}}\},  \{\underbrace{\{ \text{AP}_2 \}}_{\bm{\mathcal{C}}_{2,j}^{\text{A}}},\underbrace{ \{\text{UE}_{1}, \text{UE}_3\}\}}_{\bm{\mathcal{C}}_{2,j}^{\text{U}}} \}.
\]
Let $\mathcal{D}_{n,j}^{\text{A}}$ and $\mathcal{D}_{n,j}^{\text{U}}$ represent, respectively, the number of eAPs and UEs at the $n$-th subnetwork of the $j$-th possible configuration, where $\mathcal{D}_{n,j}^{\text{A}} = \text{Cardinality}\left\{\bm{\mathcal{C}}_{n,j}^{\text{A}}\right\}$ and $\mathcal{D}_{n,j}^{\text{U}} = \text{Cardinality}\left\{\bm{\mathcal{C}}_{n,j}^{\text{U}}\right\}, n = 1, \dots, N$ and $j = 1, \dots, \Theta\left(M,K,N\right)$. 
For this model, we assume that the $m_n$-th eAP sends a weighted sum of signals of all UEs within the $n$-th subnetwork. Accordingly, for a given cell-free network clustering configuration, $\bm{\mathcal{C}}_{j}$, the antennas of the $m$-th eAP at the $n$-th cluster (denoted by $m_n$) will have at least $\mathcal{D}_{n,j}^{\text{U}}$ streams\footnote{This assumption will require that the number of RF chains at each eAP does not fall below $\mathcal{D}_{n,j}^{\text{U}}$ \cite{RF_Chains_Specifications}.}. For simplicity, we assume that at each time instant, the $m_n$-th eAP will use only $\mathcal{D}_{n,j}^{\text{U}}\leq L$ RF chains at a time.

The NCC and the ECP will be responsible for the entire communication process. At the beginning of each time slot, the NCC will first estimate the CSI values for the UEs with respect to all serving eAPs. Then the processes of eAP clustering and per-subnetwork downlink beamforming will be performed jointly  by the  NCC and the virtual ECP.

\subsection{Channel and Antenna Model}
The communications between the eAPs and the distributed UEs occur in the 24-39 GHz mmWave bands in which transmissions suffer from limited scattering and spatial selectivity. The asymptotic orthogonality assumption among different mmWave channels does not apply to highly correlated mmWave MIMO channels \cite{mmWave_1}.
Accordingly, we adopt the well-known three-dimensional clustered model \cite{mmWave_Channel_1}. We consider a uniform planner array (UPA)\footnote{UPA is suitable for mmWave beamforming due to smaller array dimensions, ability to perform 3D beamforming (at the elevation domain), and possibility of packing many antenna elements in a small space\cite{mmWave_UPA_1,mmWave_UPA_2}.} 
at the $m_n$-th eAP and $k_n$-th UE with $a  = L_{m,1} L_{m,2}$ and $u = L_{k,1} L_{k,2}$ for which $L_{m,1}(L_{k,1})$ and $L_{m,2}(L_{k,2})$ represent the number of columns and rows of antenna elements, respectively. The downlink channel gain matrix for the $m_n \rightarrow k_n$ link (denoted by $\bm{H}_{k_n m_n} \in \mathbb{C}^{u \times a }$) can be then expressed as \cite{mmWave_Channel_1,mmWave_MIMO_Book_2}  
\begin{dmath}
    \bm{H}_{k_n m_n} = \sum_{l = 1}^{\mathcal{L}} h_{k_n m_n,l} \bm{b}_{\text{U}}\left(\vartheta_{k_n m_n,l}, \varphi_{k_n m_n,l}\right)\bm{b}^{*}_{\text{A}}\left(\theta_{k_n m_n,l}, \phi_{k_n m_n,l}\right)
    = \sum_{l = 1}^{\mathcal{L}}h_{k_n  m_n,l}\mathcal{B}\left(\bm{\theta}_{k_n  m_n,l},\bm{\phi}_{k_n m_n,l}\right),
    \label{Channel_Gain}
\end{dmath}
where $\mathcal{L}$ is the number of paths for the $m_n\rightarrow k_n$ link, $h_{k_n m_n,l}=\sqrt{\frac{1}{\kappa+\mathcal{L}-1}}\alpha_{k_n m_n,l}$ is the complex channel gain at the $l$-th path in the $m_n\rightarrow k_n$ link with $\alpha_{k_n m_n,l}\thicksim \mathcal{C N}\left(0,\sigma_{k_n m_n,l}\right)$, in which $\sigma_{k_n m_n,1} = \kappa$ (the ratio of the line-of-sight [LoS] path power to non-line-of-sight (NLoS) path power), and $\sigma_{k_n m_n,l} = 1, l = 2, \dots, \mathcal{L}$. Also, $\mathcal{B}\left(.\right) = \bm{b}_{\text{U}}\left(.\right)\bm{b}^{*}_{\text{A}}\left(.\right)$ with $\bm{\theta}_{k_n m_n} = \left[\vartheta_{k_n m_n,l},  \theta_{k_n m_n,l},\right] $ and $\bm{\phi}_{k_n m_n} = \left[\varphi_{k_n m_n,l},  \phi_{k_n m_n,l}\right]$. In (\ref{Channel_Gain}),  $\bm{b}_{\text{U}}\left(\vartheta_{k_n m_n,l}, \varphi_{k_n m_n,l}\right)\in \mathbb{C}^{u \times 1}$ and $\bm{b}_{\text{A}}\left(\theta_{k_n m_n,l}, \phi_{k_n m_n,l}\right)\in \mathbb{C}^{a \times 1}$ are the antenna array responses at the $k_n$-th UE and the $m_n$-th eAP, respectively. The antenna array response at the $m_n$-th eAP and $k_n$-th UE, respectively, can be defined as 
\begin{dmath}
    \bm{b}_{\text{A}}\left(\theta_{k_n m_n,l}, \phi_{k_n m_n,l}\right)= \left[ 
    e^{j2\pi\frac{d\left(0\sin\theta_{k_n m_n,l}\cos\phi_{k_n m_n,l}+0\sin\phi_{k_n m_n,l}\right)}{\lambda}}, \dots,
    e^{j2\pi\frac{d\left(w\sin\theta_{k_n m_n,l}\cos\phi_{k_n m_n,l}+z\sin\phi_{k_n m_n,l}\right)}{\lambda}},\\
     \dots,e^{j2\pi\frac{d\left(\left(L_{m,1}-1\right)\sin\theta_{k_n m_n,l}\cos\phi_{k_n m_n,l}+\left(L_{m,2}-1\right)\sin\phi_{k_n m_n,l}\right)}{\lambda}}
    \right]^T,
\end{dmath}
\begin{dmath}  
    \bm{b}_{\text{U}}\left(\vartheta_{k_n m_n,l}, \varphi_{k_n m_n,l}\right)= \left[ 
    e^{j2\pi\frac{d\left(0\sin\vartheta_{k_n m_n,l}\cos\varphi_{k_n m_n,l}+0\sin\varphi_{k_n m_n,l}\right)}{\lambda}}, \dots, 
    e^{j2\pi\frac{d\left(w\sin\vartheta_{k_n m_n,l}\cos\varphi_{k_n m_n,l}+z\sin\varphi_{k_n m_n,l}\right)}{\lambda}},\\ \dots,
    e^{j2\pi\frac{d\left(\left(L_{k,1}-1\right)\sin\vartheta_{k_n m_n,l}\cos\varphi_{k_n m_n,l}+\left(L_{k,2}-1\right)\sin\varphi_{k_n m_n,l}\right)}{\lambda}}
    \right]^T,
\end{dmath}
where $\theta_{k_n m_n,l}$ and $\vartheta_{k_n m_n,l}$ are the elevation angles at the $m_n$-th eAP and $k_n$-th UE, respectively, $\phi_{k_n m_n,l}$ and $\varphi_{k_n m_n,l}$ are the azimuth angles at the $m_n$-th eAP and $k_n$-th UE related to the $l$-th path in the $m_n\rightarrow k_n$ link, respectively, $d$ is the antenna spacing of eAPs and UEs, and $\lambda$ is the carrier wavelength.

\subsection{Downlink Data Transmission}
We assume that downlink transmission is performed based on two types of beamforming schemes, namely, analog RF beamsteering and baseband digital beamforming. Assuming that a certain cell-free network partitioning configuration (say $\bm{\mathcal{C}}_j$) is selected by the NCC, the received combined signal at the $i_n$-th UE can be expressed as    
\begin{dmath}
    y_{i_n} = \bm{\delta}_{i_n}^{\text{T}}\sum_{l = 1}^{N}\sum_{m_l = 1}^{\mathcal{D}_{l,j}^{\text{A}}}\bm{H}_{i_nm_l} \sum_{k_l = 1}^{\mathcal{D}_{n,j}^{\text{U}}} \bm{\mathcal{A}}_{m_l}\bm{w}_{k_lm_l}x_{k_l} + \bm{\delta}_{i_n}^{\text{T}}\bm{\eta}_{i_n}\\
    = 
    \underbrace{\sum_{m_n = 1}^{\mathcal{D}_{n,j}^{\text{A}}}
    \bm{\delta}_{i_n}^{\text{T}}
    \bm{H}_{i_nm_n}
    \bm{\mathcal{A}}_{m_n}
    \bm{w}_{i_nm_n}x_{i_n}}_{\text{Desired Signal}}
    +
    \underbrace{\sum_{m_n = 1}^{\mathcal{D}_{n,j}^{\text{A}}}
    \bm{\delta}_{i_n}^{\text{T}}
    \bm{H}_{i_nm_n}\bm{\mathcal{A}}_{m_n}
    \bm{w}_{k_n m_n}
    \sum_{k_n = 1, k_n\neq i_n}^{\mathcal{D}_{n,j}^{\text{U}}}
    x_{k_n}}_{\text{IUI}}
    + 
    \underbrace{\sum_{l = 1, l\neq n}^{N}\sum_{m_l = 1}^{\mathcal{D}_{l,j}^{\text{A}}} 
    \bm{\delta}_{i_n}^{\text{T}}
    \bm{H}_{i_nm_l}
    \sum_{k_l = 1}^{\mathcal{D}_{l,j}^{\text{U}}}
    \bm{\mathcal{A}}_{m_l}
    \bm{w}_{k_lm_l}x_{k_l}}_{\text{ISNI}}
    +   
    \underbrace{\bm{\delta}_{i_n}^{\text{T}} \bm{\eta}_{i_n}}_{\text{AWGN}},
    \label{Received_Signal_General}
\end{dmath}
where $\bm{H}_{i_nm_l}\in \mathbb{C}^{u\times a}$ is the channel gain matrix for the $m_l\rightarrow i_n$ link, $\bm{\mathcal{A}}_{m_l}\in \mathbb{C}^{a\times \mathcal{D}_{l,j}^{\text{U}}}$ is the analog RF beamsteering matrix at the $m_l$-th eAP, $\bm{w}_{k_lm_l}\in \mathbb{R}^{\mathcal{D}_{l,j}^{\text{U}}\times 1}$ is the digital baseband beamforming vector related to the $m_l\rightarrow k_l$ link, $\bm{\delta}_{m_l} \in \mathbb{C}^{u\times 1}$ is the analog beamsteering/combining vector at the $k_l$-th UE, $x_{k_l}$ is the transmitted symbol related to the $k_l$-th UE such that $\mathbb{E}\left[|x_{k_l}|^{2}\right] = P/\mathcal{D}_{l,j}^{\text{U}}$ where $P$ is the transmission power budget at each eAP, and $\bm{\eta}_{i_n}$ is the additive white Gaussian noise (AWGN) vector at the input of the $i_n$-th UE where $\bm{\eta}_{i_n}\thicksim \mathcal{N}\left(\bm{0},\sigma_{i_n}\bm{I}\right), \forall i_n = 1, 2, \dots, \mathcal{D}_{n,j}^{\text{U}}, n = 1, \dots, N$.  
The instantaneous signal-to-interference-plus-noise ratio (SINR) at the input of the $i_n$-th UE under clustering configuration $\bm{\mathcal{C}}_j$ can be expressed  as
\begin{dmath}
\gamma_{i_n}^{\left\{\bm{\mathcal{C}}_j\right\}}=
\frac
   {
    \sum_{m_n = 1}^{\mathcal{D}_{n,j}^{\text{A}}}|
    \bm{\delta}_{i_n}^{\text{T}}
    \bm{H}_{i_nm_n}
    \bm{\mathcal{A}}_{m_n}
    \bm{w}_{i_nm_n}|^2
    }
    {
    \begin{pmatrix}
    \sum_{m_n = 1}^{\mathcal{D}_{n,j}^{\text{A}}}
    \sum_{k_n = 1, k_n\neq i_n}^{\mathcal{D}_{n,j}^{\text{U}}}|
    \bm{\delta}_{i_n}^{\text{T}}
    \bm{H}_{i_nm_n}
    \bm{\mathcal{A}}_{m_n}
    \bm{w}_{k_n m_n}|^2
    + \tilde{\sigma}_{i_n}\sum_{j = 1}^u\delta_{i_nj}^2\\
    + 
    \sum_{l = 1, l\neq n}^{N}
    \left(\frac{\mathcal{D}_{n,j}^{\text{U}}}{\mathcal{D}_{l,j}^{\text{U}}}\right)^2
    \sum_{m_l = 1}^{\mathcal{D}_{l,j}^{\text{A}}} 
    \sum_{k_l = 1}^{\mathcal{D}_{l,j}^{\text{U}}}
    |
    \bm{\delta}_{i_n}^{\text{T}}
    \bm{H}_{i_nm_l}
    \bm{\mathcal{A}}_{m_l}
    \bm{w}_{k_lm_l}|^2
\end{pmatrix}
    } 
    ,
    \label{SINR_1} 
\end{dmath}
where $\tilde{\sigma}_{i_n} =\left( \frac{\sigma_{i_n}\mathcal{D}_{n,j}^{\text{U}}}{{2P}}\right)^2$.  Note that equation (\ref{SINR_1}) is derived based on the assumption that both transmitter and receiver have a full knowledge of CSI of the corresponding link. 

\subsection{General Problem Formulation} \label{Problem_Formulation_Section}
To achieve the best performance with the proposed cell-free architecture, the operations of network partitioning, analog beamsteering, and digital beamforming must be jointly optimized (e.g. by solving an optimization problem globally). The objective of this problem  can be, for example, maximization of the sum-rate of all users (i.e. {\em max-sum} objective), or maximization of the minimum rate of the users (i.e. {\em max-min} objective to achieve fairness). The general  problem formulation can be stated as follows:
\begin{equation}
\begin{aligned} 
& ~\textbf{\texttt{P}}_1: \underset{j, \left\{\bm{\mathcal{A}}_{m_n} ,\bm{\Delta}_{m_n},\bm{W}_{m_n}\right\}^{n = 1, \dots, N}_{m_n = 1, \dots, \mathcal{D}_{n,j}^{\text{A}}}}{\text{max}}~
& \text{\hspace{-43mm}} f\left(\left\{\gamma_{i_n}^{\{\bm{\mathcal{C}}_j\}}\right\}^{n = 1, \dots, N}_{i_n = 1, \dots, \mathcal{D}_{n,j}^{\text{U}}} \right)\\  
& ~\text{Subject to:} \\
&  ~  \textbf{ {C}}_1:|\bm{\mathcal{A}}_{m_n}\left(q, z\right)|^2= 1, \forall~~ q = 1, \dots, a, \;\mbox{and}\; z = 1, \dots, \mathcal{D}_{n,j}^{\text{U}},
\\
&  ~  \textbf{ {C}}_2:|\bm{\delta}_{k_n}\left(q\right)|^2 = 1, \forall~~k_n = 1, \dots, \mathcal{D}_{n,j}^{\text{U}}, \;\mbox{and}\; q = 1, \dots, u, 
\\
&  ~  \textbf{ {C}}_3:||\bm{W}_{m_n}\left(\left[1 \dots \mathcal{D}_{n,j}^{\text{U}} \right],z\right)||^2\leq 1, \forall~~  z = 1, \dots, \mathcal{D}_{n,j}^{\text{U}},
\end{aligned}\label{Opt_Prob_1}
\end{equation}
where  $\bm{\mathcal{A}}_{m_n}\in \mathbb{C}^{a \times \mathcal{D}_{n,j}^{\text{U}}} $ is the analog beamsteering matrix at the $m_n$-th eAP,
 $\bm{W}_{m_n}\in \mathbb{R}^{\mathcal{D}_{n,j}^{\text{U}}\times \mathcal{D}_{n,j}^{\text{U}}}$
is the digital beamforming matrix at the $m_n$-th eAP  with $\bm{W}_{m_n} = \left[\bm{w}_{1_nm_n}, \dots, \bm{w}_{\mathcal{D}_{n,j}^{\text{U}}m_n} \right]$, where $\bm{w}_{k_n m_n} = \left[\sqrt{w_{k_n m_n,1}}, \dots, \sqrt{w_{k_n m_n,\mathcal{D}_{n,j}^{\text{U}}}} \right]^{\text{T}}$ is the digital baseband beamforming vector related to the $m_n\rightarrow k_n$ link, 
 $\bm{\Delta}_{m_n}\in \mathbb{R}^{u\times \mathcal{D}_{n,j}^{\text{U}}}$
is the analog beamsteering/combining matrix related to all UEs of the $m_n$-th subnetwork, in which $\bm{\Delta}_{m_n} = \left[\bm{\delta}_{1_n}, \dots, \bm{\delta}_{\mathcal{D}_{n,j}^{\text{U}}} \right]$, where $\bm{\delta}_{k_n} = \left[\delta_{k_n 1}, \dots, \delta_{k_n u} \right]^{\text{T}}$ is the analog beamsteering/combining vector at the $k_n$-th UE. Furthermore, the index $j\in \left[1,\dots, \Theta(M,K,N)\right]$ refers to the selected cell-free partitioning configuration. 

 $\textbf{\texttt{P}}_1$ is a combinatorial optimization problem which is characterized by: i) non-convexity of the objective function $f\left(.\right)$ (discrete $j$), ii) discrete nature of optimization variable $j$, iii) non-affine nature of the constraints $\textbf{ {C}}_1$ and $\textbf{ {C}}_2$. 
To solve  $\textbf{\texttt{P}}_1$ optimally, a simultaneous optimization for $j$, $\bm{\mathcal{A}}_{m_n}$,  $\bm{\Delta}_{m_n}$, and 
$\bm{W}_{m_n},~\forall~n = 1, \dots, N$, $m_n = 1, \dots, \mathcal{D}_{n,j}^{\text{A}}$, and $k_n = 1, \dots, \mathcal{D}_{n,j}^{\text{U}}$ will be required. This is achieved by going through every possible clustering configuration of the cell-free network $(\bm{\mathcal{C}}_j, j =  1, \dots, \Theta\left(M, K, N\right))$, and for each $\bm{\mathcal{C}}_j$, we will need to find the corresponding optimal analog beamsteering and digital beamforming matrices (i.e. $\bm{\mathcal{A}}_{m_n}$,  $\bm{\Delta}_{m_n}$, and 
$\bm{W}_{m_n}$). \textit{The globally optimal solution is then the one that gives the best performance among all possible  clustering configurations and the corresponding matrices $\bm{\mathcal{A}}_{m_n}$,  $\bm{\Delta}_{m_n}$, and 
$\bm{W}_{m_n}$.} The solution will have a combinatorial computational complexity in terms of the network parameters such as $M, K, N, a_{m_n}$, and $u_{k_n}$ (see Section \ref{Complexity} for further discussions). 

\section{mmWave Hybrid Beamforming Design}
As has been mentioned in the last section, the problem $\textbf{\texttt{P}}_1$ in (\ref{Opt_Prob_1}) is a combinatorial optimization problem with four overlapping feasible spaces (spaces of $j$, $\bm{\mathcal{A}}_{m_n}$,  $\bm{\Delta}_{m_n}$, and 
$\bm{W}_{m_n}$). Such a problem can be solved by global optimization techniques such as deterministic methods (e.g. inner and outer approximation and cutting-plane methods), stochastic methods (e.g. direct Monte-Carlo sampling and stochastic tunneling), and heuristic methods (e.g. genetic algorithms and swarm-based optimization algorithms) \cite{Boyd2004}. However, generating an efficient solution of $\textbf{\texttt{P}}_1$ with reasonable computational complexity and short computing time becomes very challenging as the number of eAPs and/or UEs increases. 
In this section, we develop an efficient low-complexity mixed DRL-cum-convex optimization-based solution of  $\textbf{\texttt{P}}_1$. In the following, we will first discuss the problem of downlink beamforming at each cell-free subnetwork  and then develop a hierarchical DRL-based scheme that jointly performs network clustering and per-subnetwork hybrid beamforming. 
\subsection{Hybrid Beamforming for Cell-Free MIMO} \label{Hybrid_Beamforming_Section}
\begin{figure}[htb]
		\centering
		\includegraphics[scale = 0.185]{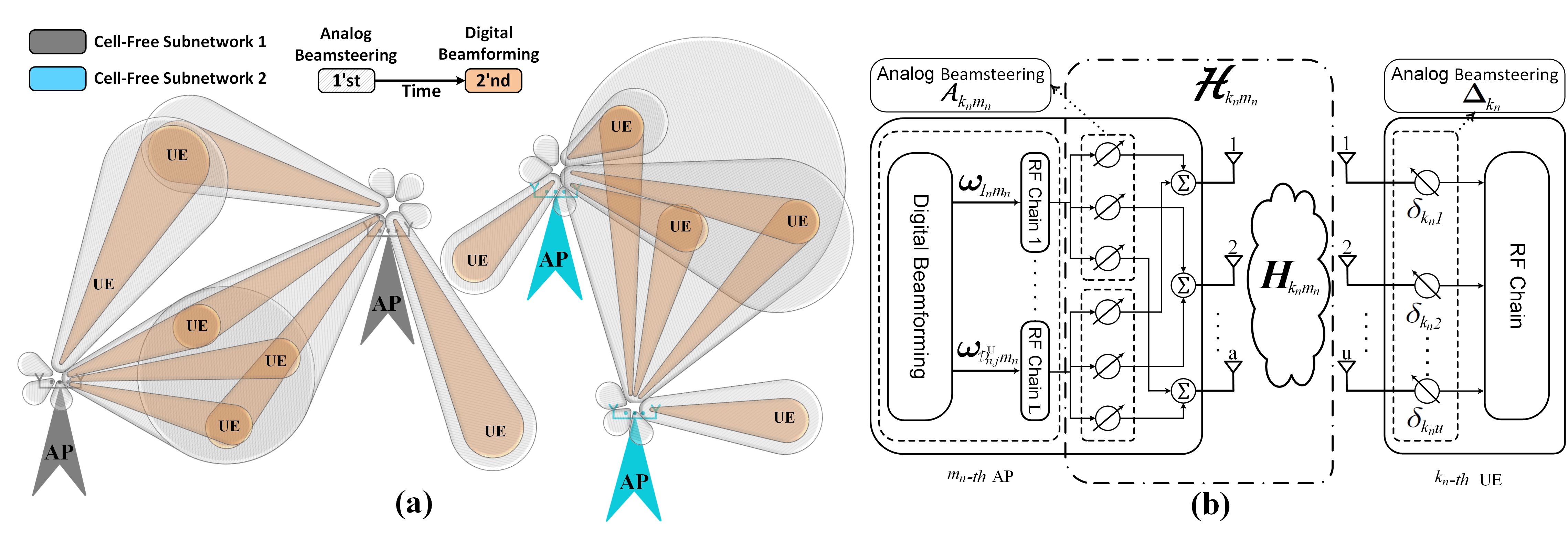}
		\caption{Hybrid Beamforming: (a) Example scenario, (b) Block diagram.}\label{Beamforming_Model}
	\end{figure}

Partitioning the overall cell-free network architecture (i.e. eAPs and UEs) into a set of computationally independent cell-free subnetworks introduces ISNI to the received signal. Furthermore, simultaneous in-band transmission will cause IUI to all UEs that belong to the same subnetwork. We develop a novel hybrid analog-digital beamforming scheme that efficiently mitigates the effects  of ISNI and IUI. Specifically, in the proposed method, downlink beamforming at the multi-antenna eAPs within each cluster is performed in two consecutive stages (\textbf{Fig. \ref{Beamforming_Model}(a)}).
Under a certain network partitioning configuration (e.g. $\bm{\mathcal{C}}_j$ for some $j$),  each cell-free subnetwork first performs an analog beamsteering for all eAPs such that the ISNI from the nearby clusters is minimized. This is achieved by directing the main beams of eAPs (i.e. the main lobes of antennas at each eAP) to the UEs belonging to the same cluster and setting the beam directions of annihilated side lobes to the UEs located outside the intended subnetwork (ISNI minimization). Once the beams of different subnetworks are steered to their desired coverage areas, digital beamforming (\textbf{Fig. \ref{Beamforming_Model}(b)})  is performed at each eAP to maximize the overall performance and mitigate IUI for the UEs that are located within the intended cell-free subnetwork. Digital beamforming is performed using the overall effective channel after applying the analog beamsteering phase matrices to the original CSI matrices. 

For the proposed beamforming scheme, each eAP is assumed to be equipped with $L$ RF chains. Each UE within each subnetwork is assigned to one communication stream by each eAP in that subnetwork\footnote{A single baseband communication stream is handled by a single RF chain.}. To achieve this, the number of UEs at each cell-free subnetwork must not exceed the number of RF chains at each eAP (i.e. $\mathcal{D}_{n,j}^{\text{U}}\leq L$). \textbf{Fig. \ref{Beamforming_Model}(b)} illustrates the functional block diagram of the $m_n$-th eAP transmitter/beamformer and the $k_n$-th UE.
At the UE side, we assume that signal from different antennas are combined through a low-complexity analog beamsteering/combining scheme using the analog combining vector $\bm{\delta}_{k_n}$. After analog beamsteering at all eAPs and UEs,  digital beamforming takes place considering the effective CSI obtained after applying analog beamsteering at both eAPs and UEs (i.e. after applying $\bm{\mathcal{A}}_{m_n}$ and $\bm{\delta}_{k_n}, ~\forall~n,m,k$).
\subsection{Analog Beamsteering Subsystem: ISNI Mitigation}

The main beam (also referred to as the main lobe) of an antenna element contains the largest portion of the field strength (either radiated or absorbed). The direction of the main beam of a single antenna can be adjusted to match the direction of arrival of the transmitted signal (elevation and azimuth angles) \cite{Balanis-2016-antenna}. We propose an analog beamsteering technique to be used in the first-stage of downlink signal transmission at each cell-free subnetwork. This is achieved by utilizing the spatial signatures between UEs of the overall cell-free network and those of eAPs of the intended subnetwork. Specifically, analog beamsteering in a cell-free subnetwork is used to minimize the ISNI  caused to the UEs from outside the intended subnetwork. This is achieved by directing the main lobes of the eAPs of each subnetwork toward the UEs belonging to the same subnetwork and setting the directions of the weakest beam side lobes to those outside the intended subnetwork.\\
\textbf{{Designing the beamsteering objective function:}} In order to mitigate the ISNI components, analog beamsteering matrices $\bm{\mathcal{A}}_{m_n}, \forall~ m_n = 1, \dots, \mathcal{D}_{n,j}^{\text{U}}$ and $n = 1, \dots, N$ at the $m_n$-th eAP have to be designed such that they ``\textit{zero-force}" the communication links between all eAPs of the $n$-th subnetwork with UEs outside the intended subnetwork. At the same time, the communication links between all eAPs of the $n$-th subnetwork and UEs inside the intended subnetwork are optimized. To achieve this, let us first define the ``\textit{null space}" of an arbitrary $m_x\rightarrow k_y$ MIMO link using the following axiom.
\begin{axiom}\label{axiom_one} 
Let $\bm{H}_{k_ym_x}\in \mathbb{C}^{u \times a}$ be an arbitrary mmwave MIMO channel matrix. If the singular value decomposition (SVD) of $\bm{H}_{k_ym_x}$ is given by   
\begin{dmath}
    \bm{H}_{k_ym_x} = \bm{U}_{k_ym_x}\bm{\Sigma}_{k_ym_x} \bm{V}_{k_ym_x}^{*}
     =  \left[\bm{U}_{k_ym_x}^{(1)}\bm{U}_{k_ym_x}^{(0)}\right]\bm{\Sigma}_{k_ym_x}\left[\bm{V}_{k_ym_x}^{(1)}, \bm{V}_{k_ym_x}^{(0)}\right]^{*},
\end{dmath}
then the left null space of $\bm{H}_{k_ym_x}$ is given by
\begin{dmath}
    \text{Null}_{\textit{L}}\left(\bm{H}_{k_ym_x}\right) = \bm{U}_{k_ym_x}^{(0)}.
\end{dmath}
Furthermore, the right null space of $\bm{H_{k_ym_x}}$ is given by
\begin{dmath}
    \text{Null}_{\textit{R}}\left(\bm{H}_{k_ym_x}\right) = \bm{V}_{k_ym_x}^{(0)},
\end{dmath}
where $\bm{U}_{k_ym_x}\in \mathbb{C}^{u \times u }$ and $\bm{V}_{k_ym_y}\in \mathbb{C}^{a \times a }$ are unitary matrices, and $\bm{\Sigma}_{k_ym_x}\in \mathbb{R}^{u \times a }$ is a diagonal matrix containing the eigenvalues of $\bm{H}_{k_ym_x}$. $\bm{U}_{k_ym_x}^{(1)}\in \mathbb{C}^{u \times r}$ and $\bm{V}_{k_ym_x}^{(1)}\in \mathbb{C}^{a \times r}$ are the matrices with columns from $\bm{U}_{k_ym_x}$ and $\bm{V}_{k_ym_x}$, respectively, corresponding to the non-zero eigenvalues of $\bm{H}_{k_ym_x}$,  and $\bm{U}_{k_ym_x}^{(0)}\in  \mathbb{C}^{u \times \left(u - r\right)}$  $\bm{V}_{k_ym_x}^{(0)}\in  \mathbb{C}^{a \times \left(a - r\right)}$ with columns from $\bm{U}_{k_ym_x}$ and $\bm{V}_{k_ym_x}$, respectively, corresponding to the zero eigenvalues of $\bm{H}_{k_ym_x}$,  where $r = \text{rank}\left(\bm{H}_{k_ym_x}\right)$. 
\end{axiom}  

\noindent
\textbf{Remark:} To guarantee the existence of a null space for any arbitrary $\bm{H}_{k_ym_x}$, the number of antennas at each eAP must exceed that of UEs served by that eAP. This condition complies with the fact that mmWave networks use massive MIMO systems at all distributed eAPs. Given the left and right null spaces of $\bm{H}_{k_ym_x}$, 
the projection of a complex vector $\bm{\delta}_{k_y}$ into $\bm{U}_{k_ym_x}^{(0)}$ can be given by
\begin{dmath}
    \bm{\delta}_{k_ym_x}^{\perp} = \bm{\delta}_{ky}^{\text{T}}\bm{U}_{k_ym_x}^{(0)}\left( \bm{U}_{k_ym_x}^{(0)} \right)^{*}. \label{Projection_1}
\end{dmath}
Furthermore, the projection of a matrix $\bm{\mathcal{A}}_{m_x}$ into $\bm{V}_{k_ym_x}^{(0)}$ can be given by  
\begin{dmath}
    \bm{\mathcal{A}}_{k_ym_x}^{\perp} = \bm{V}_{k_ym_x}^{(0)}\left( \bm{V}_{k_ym_x}^{(0)} \right)^{*}\bm{\mathcal{A}}_{m_x}.\label{Projection_2}
\end{dmath}
 (\ref{Projection_1}) and (\ref{Projection_2}) above can be derived using the fact that $\left(\bm{\delta}_{k_y} - \bm{\delta}_{k_ym_x}^{ \perp}  \right)^{\text{T}} \bm{U}_{k_ym_x}^{(0)} = \bm{0}$, $\left( \bm{V}_{k_ym_x}^{(0)} \right)^{*}\left(\bm{\mathcal{A}}_{m_x} - \bm{\mathcal{A}}_{k_ym_x}^{\perp}  \right) = \bm{0}$ and 
 $\bm{U}_{k_ym_x}^{(0)}\left( \bm{U}_{k_ym_x}^{(0)} \right)^{*} = \bm{I}_{u}$, $\bm{V}_{k_ym_x}^{(0)}\left( \bm{V}_{k_ym_x}^{(0)} \right)^{*} = \bm{I}_{a}$, where $\bm{I}_{u}\in \mathbb{R}^{u\times u}$, $\bm{I}_{a}\in \mathbb{R}^{a \times a}$ are identity matrices. Using a similar procedure, the projection of $\bm{\delta}_{k_y}$ and  $\bm{\mathcal{A}}_{m_x}$ on the left and right ``\textit{non-annihilating}" subspaces of $\bm{H}_{k_ym_x}$ can be given, respectively, by \footnote{Here, the non-annihilating subspace refers to the subspace of $\bm{H}_{k_ym_x}$ after  it's null subspace has been removed.}
 \begin{dmath}
    \bm{\delta}_{k_ym_x}^{\not \perp} = \bm{\delta}_{k_y}^{\text{T}}
    \bm{U}_{k_ym_x}^{(1)}\left( \bm{U}_{k_ym_x}^{(1)} \right)^{*}, \; \mbox{and}\label{Projection_21}
\end{dmath}
\begin{dmath}
    \bm{\mathcal{A}}_{k_ym_x}^{\not \perp} = \bm{V}_{k_ym_x}^{(1)}\left( \bm{V}_{k_ym_x}^{(1)} \right)^{*}\bm{\mathcal{A}}_{m_x}.\label{Projection_22}
\end{dmath}
The objective of analog beamsteering is to reduce ISNI within each cell-free subnetwork. However, focusing only on ZF technique to remove the interfering beams (i.e. ISNI) between adjacent subnetworks may result in misalignment of the main beams of the eAP antennas with those of the UEs within the same subnetwork, and hence reduced/inappropriate antenna directivity. Therefore, we propose a novel analog beamsteering scheme based on maximizing the so-called ``\textit{secrecy sum power gains}" at each subnetwork. This is done at the $n$-th subnetwork by maximizing the sum of powers of two channel projections: i) the non-annihilating projections of $\bm{\delta}_{k_n}$ and $\bm{\mathcal{A}}_{m_n}$ on $\bm{H}_{k_n m_n}$ (i.e, $\bm{\delta}_{k_n m_n}^{\not \perp}$ and $\bm{\mathcal{A}}_{k_n m_n}^{\not \perp},~\forall~k_n = 1, \dots, \mathcal{D}_{n,j}^{\text{U}} ~\&~~m_n = 1, \dots, \mathcal{D}_{n,j}^{\text{A}}$), and ii) the annihilating projections of $\bm{\mathcal{A}}_{m_n}$ on $\bm{H}_{k_lm_n}$ (i.e.  $\bm{\mathcal{A}}_{k_lm_n}^{ \perp},~ \forall~ l\neq n$). 

We are now ready to formulate the beamsteering optimization problem $\textbf{\texttt{P}}_2$ to mitigate ISNI, which will need to be solved for the $n$-th cell-free subnetwork,  as follows:
\begin{equation}
\begin{aligned}
& ~\textbf{\texttt{P}}_2: \underset{\left\{\bm{\delta}_{k_n}, \bm{\mathcal{A}}_{m_n}\right\}^{m_n =1, \dots, \mathcal{D}_{n,j}^{\text{A}}}_{k_n = 1, \dots, \mathcal{D}_{n,j}^{\text{U}}}}{\text{max}}~
& \text{\hspace{-82mm}} \sum_{m_n = 1}^{\mathcal{D}_{n,j}^{\text{A}}}\left(\sum_{k_n = 1}^{\mathcal{D}_{n,j}^{\text{U}}}
    ||\bm{\delta}^{\not \perp}_{k_n m_n}\bm{\Sigma}_{k_n m_n}\bm{\mathcal{A}}_{k_n m_n}^{\not \perp}||^2\right.
    \\
&  \text{\hspace{45mm}}  \left.+ \sum_{l = 1, l \neq n}^{N}\sum_{k_l = 1}^{\mathcal{D}_{l,j}^{\text{U}}}
    || \bm{\delta}^{  \perp}_{k_lm_n}\left(t-1\right)\bm{\Sigma}_{k_lm_n}\bm{\mathcal{A}}_{k_lm_n}^{\perp}||^2
    \right),\\
& ~\text{Subject to:} \\
&  ~  \textbf{{C}}_1:|\bm{\mathcal{A}}_{m_n}\left(q, z\right)|^2= 1, \forall~q = 1, \dots, a \;\mbox{and}\; z = 1, \dots, u, 
\\
&  ~  \textbf{{C}}_2:|\bm{\bm{\delta}}_{k_n}\left(q\right)|^2= 1, \forall~~ q = 1, \dots, u.
\end{aligned}\label{Opt_Prob_2}
\end{equation}
In the objective function of  $\textbf{\texttt{P}}_2$ in (\ref{Opt_Prob_2}), the variable matrices $\bm{\mathcal{A}}_{k_n m_n}^{\not \perp}$, $\bm{\mathcal{A}}_{k_n m_n}^{\not \perp}$, and $\bm{\delta}^{\not \perp}_{k_n m_n}$ are non-linear functions of $\bm{\delta}_{k_y}$ and $\bm{\mathcal{A}}_{m_n}$. This relationship can be inferred from (\ref{Projection_21}) and (\ref{Projection_22}) as a non-linear truncation of unitary matrices of the SVD of $\bm{\delta}_{k_y}$ and $\bm{\mathcal{A}}_{m_n}$. Accordingly,  $\textbf{\texttt{P}}_2$ is a non-convex combinatorial optimization problem.

\subsection{Digital Beamforming Subsystem: Transmission Rate Maximization}
After the analog beamsteering at all MIMO transmitters (eAPs) and receivers (UEs) has been performed, the actual CSI matrices (i.e. $\bm{H}_{k_n m_n}, ~\forall~n, m_n$, and $k_n$) will be multiplied by the beamsteering matrices (from the right side) and the analog combining/beamsteering vectors (from the left side). Accordingly, the effective channel gain at the $m_n\rightarrow k_n$ link (denoted by $\bm{\mathcal{H}}_{k_n m_n}\in \mathbb{C}^{1 \times \mathcal{D}_{n,j}^{\text{U}}}$) will be given by  
\begin{dmath}
\bm{\mathcal{H}}_{k_n m_n} =  \bm{\delta}_{k_n}^{\text{T}}\bm{H}_{k_n m_n}\bm{\mathcal{A}}_{m_n}.
\end{dmath}
Note that, the $i_n$-th element in  $\bm{\mathcal{H}}_{k_n m_n}$ ($i_n = 1_n, \dots, \mathcal{D}_{n,j}^{\text{U}}$) corresponds to the signal radiated from the beam steered at UE $i_n$ within the $m_n$-th subnetwork. However, each element in $\bm{\mathcal{H}}_{k_n m_n}$ will contain portions of signals send to all UEs of the $n$-th subnetwork.
\\
\textbf{UE ordering and SIC decoding:} In downlink multiuser single-input single-output (SISO) wireless networks, SIC-based UEs (usually referred to as non-orthogonal multiple access [NOMA] UEs) are first ordered based on their instantaneous channel gains. Then, the UEs with lower link gains are allocated higher transmission power compared to those with better communication link. At the receiver side, multi-level SIC operations are conducted such that the interfering signals related to the UEs with lower channel gains are decoded and then subtracted \cite{NOMA_Principles}. When the UEs of a downlink NOMA system are served by a single eAP with multiple antennas or by multiple single-antenna APs, the UEs can be ordered based on the norm of their channel quality vector \cite{GCoMP}. Such a channel quality metric is denoted as the ``effective channel gain". In our considered network model, however, all the eAPs and the UEs in each cluster are assumed to be equipped with multiple antennas. Therefore, the channel gain between each eAP and any arbitrary NOMA UE is represented by a complex matrix. We use the squared norm of the effective channel gain vectors (i.e. $\bm{\mathcal{H}}_{k_n m_n}$) as the NOMA effective channel gain. Accordingly, we assume that the UEs within the $n$-th cluster are arranged in an ascending order as follows:
\vspace{0.3cm}

$\frac{\sum_{m_n = 1}^{\mathcal{D}_{n,j}^{\text{A}}}|
    \bm{\mathcal{H}}_{1_nm_n}|^2}{
    \sum_{l = 1, l\neq n}^{N}\sum_{m_l = 1}^{\mathcal{D}_l^{\text{A}}} \sum_{k_l = 1}^{\mathcal{D}_l^{\text{U}}}
    |\bm{\mathcal{H}}_{k_lm_l}|^2} 
    \leq 
    \dots
    \leq
    \frac{\sum_{m_n = 1}^{\mathcal{D}_{n,j}^{\text{A}}}|
    \bm{\mathcal{H}}_{i_nm_n}|^2}{
    \sum_{l = 1, l\neq n}^{N}\sum_{m_l = 1}^{\mathcal{D}_l^{\text{A}}} \sum_{k_l = 1}^{\mathcal{D}_l^{\text{U}}}
    |\bm{\mathcal{H}}_{k_lm_l}|^2}
    \leq
    \dots
    \leq
    \frac{\sum_{m_n = 1}^{\mathcal{D}_{n,j}^{\text{A}}}|
    \bm{\mathcal{H}}_{\mathcal{D}_{n,j}^{\text{U}}m_n}|^2}{
    \sum_{l = 1, l\neq n}^{N}\sum_{m_l = 1}^{\mathcal{D}_l^{\text{A}}} \sum_{k_l = 1}^{\mathcal{D}_l^{\text{U}}}
    |\bm{\mathcal{H}}_{k_lm_l}|^2}$.
    
\vspace{0.4cm}    
Note that, for the UE ordering, we divide the effective channel metric of each UE by the corresponding ISNI assuming the availability of channel gain matrices of the overall cell-free network at each eAP. With proper  power allocation for the UEs, at the receiver side, the $i_n$-th UE located in the $n$-th cluster will be able to remove interference components from $i_n-1$ UEs with higher overall power gain.
Accordingly, $\gamma_{i_n}^{\left\{\bm{\mathcal{C}}_j\right\}}$ can be rewritten as
\begin{dmath}
\gamma_{i_n}^{\left\{\bm{\mathcal{C}}_j\right\}}=
\frac
   {
    \sum_{m_n = 1}^{\mathcal{D}_{n,j}^{\text{A}}}|\bm{\mathcal{H}}_{i_nm_n}
    \bm{w}_{i_nm_n}|^2
    }
    {
    \begin{pmatrix}
    \sum_{m_n = 1}^{\mathcal{D}_{n,j}^{\text{A}}}
    \sum_{k_n = i_n + 1}^{\mathcal{D}_{n,j}^{\text{U}}}|\bm{\mathcal{H}}_{i_nm_n}
    \bm{w}_{k_n m_n}|^2
    + \tilde{\sigma}_{i_n}\sum_{j = 1}^u\delta_{i_nj}^2\\
    + 
    \sum_{l = 1, l\neq n}^{N}
    \left(\frac{\mathcal{D}_{n,j}^{\text{U}}}{\mathcal{D}_{l,j}^{\text{U}}}\right)^2
    \sum_{m_l = 1}^{\mathcal{D}_{l,j}^{\text{A}}} 
    \sum_{k_l = 1}^{\mathcal{D}_{l,j}^{\text{U}}}
    |
    \bm{\mathcal{H}}_{i_nm_l}
    \bm{w}_{k_lm_l}|^2
\end{pmatrix}
    } 
    ,
    \label{SINR_2} 
\end{dmath}
where $\tilde{\sigma}_{i_n} =\left( \frac{\sigma_{i_n}\mathcal{D}_{n,j}^{\text{U}}}{{2P}}\right)^2$. 


We propose a beamforming (i.e. precoding) scheme that maximizes the sum-rate of the UEs. The digital beamforming problem at the $n$-th cell-free subnetwork under a certain cell-free network partitioning configuration ($\bm{\mathcal{C}}_j$) can be formulated as
\begin{equation}
\begin{aligned}
& ~\textbf{\texttt{P}}_3: \underset{\left\{\bm{w}_{k_n m_n}\right\}_{k_n = 1, \dots, \mathcal{D}_{n,j}^{\text{U}}}^{m_n = 1, \dots, \mathcal{D}_{n,j}^{\text{A}}}}{\text{max}}~
& \text{\hspace{-65mm}} \sum_{n = 1}^{N}\sum_{i_n=1}^{\mathcal{D}_{n,j}^{\text{U}}}\log_2\left(1+\gamma_{i_n}^{\{\bm{\mathcal{C}}_j\}} \right)\\
& ~\text{Subject to:} \\
& ~\textbf{{C}}_1: \sum_{m_n = 1}^{\mathcal{D}_{n,j}^{\text{A}}} \left(||
    \bm{\mathcal{H}}_{i_nm_n}\bm{w}_{\delta_{i_n}m_n}||^2
-
\sum_{w = \delta_{i_n} + 1}^{i_n}
||    \bm{\mathcal{H}}_{i_nm_n}\bm{w}_{wm_n}||^2 \right)\geq \epsilon, \\
&  ~  \textbf{{C}}_2:||\bm{w}_{k_n m_n}||^2\leq 1,
\\
& \text{\hspace{10mm}} \forall~\delta_{i_n}= 1, \dots, {i_n}-1, l=2, \dots, \mathcal{D}_{n,j}^{\text{U}}, m_n = 1, \dots, \mathcal{D}_{n,j}^{\text{A}},\\
& \text{\hspace{10mm}} k_n = 1, \dots, \mathcal{D}_{n,j}^{\text{U}},  \;\mbox{and}\;  n = 1, \dots, N. \\
\end{aligned}\label{Opt_Prob_3}
\end{equation}
In \cite[Appendix C]{GCoMP}, it was shown that problem $\textbf{\texttt{P}}_3$ in (\ref{Opt_Prob_3}) represents a  convex optimization problem under the assumption of UE ordering and SIC-based decoding. Specifically, it was shown that the objective function of $\textbf{\texttt{P}}_3$ in (\ref{Opt_Prob_3}) can be decomposed into a sum of convex and concave functions with the convex function having a more increasing rate than that of the concave one. Furthermore, it is easy to confirm that the constraints $\textbf{{C}}_1$ and $\textbf{{C}}_2$ represent affine relations of $\bm{w}_{k_n m_n}$\footnote{This can be easily confirmed by rewriting the vector form of $\textbf{{C}}_1$ and $\textbf{{C}}_2$ in a sum of products format, rather than vector format.}. This convex problem can be easily solved by using the Karush-Kuhn-Tucker (KKT) conditions and utilizing some numerical methods for calculating the first and second differentiation of the Lagrangian function\footnote{Despite the existence of the second derivative of the objective function, a closed-form expression is difficult to derive due to the multi-dimensional nature of the optimization variables.}.

\section{Hierarchical DRL Design: Joint Network Partitioning and Hybrid Beamforming}
  
\subsection{DRL Techniques for Solving Optimization Problems}

DRL techniques have been used to solve optimization problems in wireless communications systems (e.g. for optimization of downlink power control in a multi-cell system~\cite{Ishfaq}, beamforming optimization in a cell-free network~\cite{Dynamic_Cell_Free}).  In these cases, a DRL agent (e.g. a network entity) aims at learning  the ``optimal" mapping between a system state $\bm{s}$ and the action $\bm{a}$ (e.g. a policy function or a value function) in order to maximize its reward discounted reward over a time horizon. Depending on the agent's objective, DRL  techniques are commonly classified into three categories:
\begin{itemize}
    \item \textit{Value-based} methods such as deep Q-learning (DQL) and state–action–reward–state–action (SARSA) learn the value function $V(\bm{s})$ or the state-value function $Q(\bm{s}, \bm{a} )$ to find a policy.
    \item \textit{Policy-based} methods obtain the mapping between the system state and the action (i.e. policy) directly. These methods generally suffer from noisy gradients and high variance~\cite{RL_Book}.
    \item \textit{Actor-critic} methods are a hybrid of the value-based and policy-based methods. Value-based methods are used to reduce the variance of the policy-based methods by estimating the value function or the action-value function (a.k.a. the critic) to improve the performance of the policy (a.k.a. the actor).
\end{itemize}
 
\subsection{Hierarchical DRL Architecture}
\begin{figure}[htb]
		\centering
		\includegraphics[scale = 0.27]{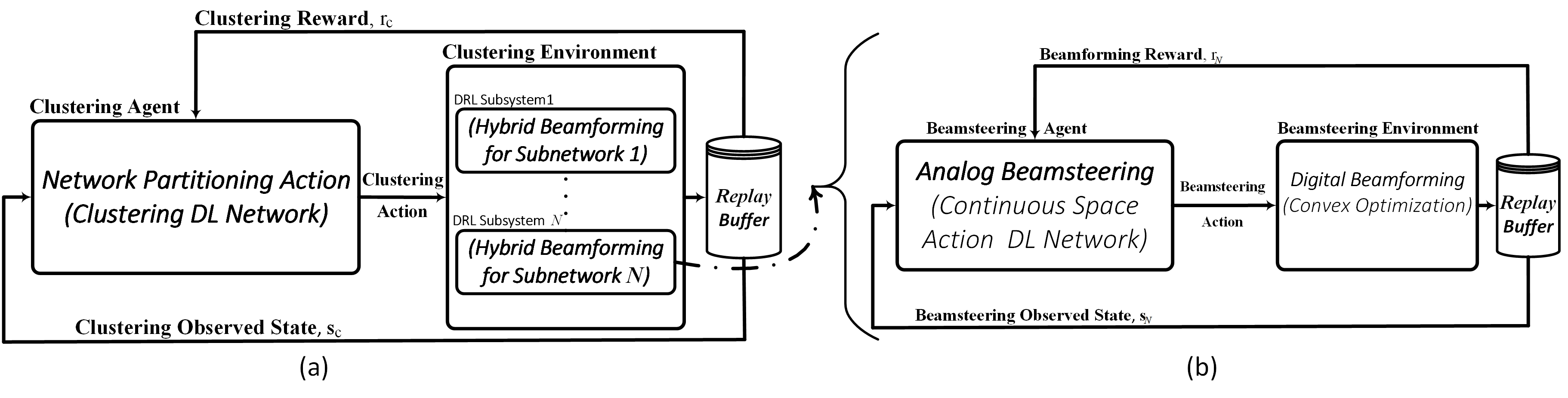}
		\caption{Block diagram of the hierarchical DRL clustering system.}\label{DRL_1}
\end{figure}

The proposed solution consists of two-levels of interacting DRL models. The first-level of the proposed system is responsible for network partitioning (i.e. clustering) and it consists of a single DRL model. The agent of the first-level DRL model is located at the NCC and is mainly responsible for partitioning the overall cell-free network into a set of non-overlapping cell-free subnetworks. The second-level of the proposed hierarchical architecture consists of $N$ independent DRL subsystems. Each DRL subsystem is responsible for conducting the hybrid analog beamsteering-digital beamforming process in a single cell-free subnetwork. { This is achieved by training the DRL subsystem agent to optimize the analog beamsteering vectors of all eAPs and UEs of the corresponding cell-free subnetwork while the digital beamforming problem for the same subnetwork is modeled and solved as a convex optimization problem inside the environment of the DRL subsystem for analog beamsteering. All of the DRL subsystems are within the environment of the first-level DRL clustering system} (\textbf{Fig. \ref{DRL_1}(a)}). 
In terms of the time-scale of operation, the overall cell-free network is assumed to cluster (partition) every $\tau \geq 1$ time instants\footnote{The value of $\tau$ may be considered as a design parameter that can depend, for example, on the time-varying nature of the propagation environment.}. On the other hand, the hybrid beamforming process is assumed to take place at each time slot. Further details about the DRL model action spaces, rewards, and observed states are described in the following sections.

\subsection{Learn to ``\textit{Cluster}"}\label{Learn_to_Cluster}

Our objective is to design a self-organizing cell-free network that has the ability to self-partition (self-cluster) into a group of cell-free subnetworks, in a time-slot basis, based on the instantaneous CSI. For the proposed dynamic cell-free network with $M$ eAPs, $K$ UEs, and $N$ subnetworks,  there will be $\Theta\left(M,N \right) = \left(\frac{N!}{\sqrt{2}}\right)^2C\left(M,N\right)C\left(K,N\right)$ possible configurations for cell-free subnetworks, where $C\left(M,N\right)$ is the Stirling number, which can be calculated as 
\cite{BellNumber}
\begin{equation}\label{eq:bell-number}
    C\left(M,N\right)=
    \left\{
	\begin{array}{ll}
		M \\
		N
	\end{array}
\right\}
=\frac{1}{N!}
\sum_{i = 0}^{N}\left(-1\right)^{i}\binom{N}{i}\left(N-i\right)^M.
\end{equation}
Optimally updating the cell-free network configuration on a time-slot basis requires going through all possible configurations (as discussed in Section \ref{Problem_Formulation_Section}) which will be practically infeasible for a massive cell-free network with large numbers of eAPs and UEs\footnote{As an example, for $M = 100$, $K = 50$, and $N = 10$, there will be approximately $1.28962122\times10^{138}$ possible cell-free subnetwork configurations.}. 

In this section, we design several low-complexity DRL-based methods to efficiently perform network clustering on a time-slot basis. Each of these methods accepts  certain network information (e.g. instantaneous CSI values across the entire network) and outputs a certain network partitioning configuration that maximizes a predefined network performance metric.
\textbf{Table} \ref{Table3} summarizes the environment design for the DRL models in terms of the problem parameters.
\begin{table}[h!]
    \centering
      \caption{DRL model for network partitioning}
\begin{tabular}{|c|c|}
\hline
\shortstack{Clustering Environment\\ Variables} & \shortstack{Network Partitioning System Equivalence\\  \textcolor{white}{.}} \\
\hline
\hline
State ${\bm{s}_{\text{c}}}=\left\{{s}_{\text{c},1}, \dots, {s}_{\text{c},N} \right\}$ & $\left\{\prod_{t = 1}^{\tau}\prod_{i_1 =1}^{\mathcal{D}_{1,j}^{\text{U}}}\gamma_{i_1}^{\{\mathcal{C}_j\}}\left(t\right), \dots, 
\prod_{t = 1}^{\tau}\prod_{i_N =1}^{\mathcal{D}_{N,j}^{\text{U}}}\gamma_{i_N}^{\{\mathcal{C}_j\}}\left(t\right)\right\}$\\
\hline
Reward $r_{\text{c}}$ & $ \prod_{t = 1}^{\tau}\left(\prod_{n = 1}^N\left(\sum_{i_n = 1}^{\mathcal{D}_{n,j}^{\text{U}}}  \;\log\left(1+\gamma_{i_n}^{\{\mathcal{C}_j\}}\left(t\right)\right)\right)\right)$\\
\hline
Action $\bm{a}_{\text{c}}$ & $\bm{\mathcal{C}}_j = \left\{\left\{ \bm{\mathcal{C}}_{1,j}^{\text{A}},\bm{\mathcal{C}}_{1,j}^{\text{U}} \right\}, \dots, \left\{ \bm{\mathcal{C}}_{N,j}^{\text{A}},\bm{\mathcal{C}}_{N,j}^{\text{U}} \right\} \right\}$\\
\hline
\end{tabular}
    \label{Table3}
\end{table}
Note that, in \textbf{Table} \ref{Table3}, the DRL system state vector ($\bm{s}_{\text{c}}$) corresponds to a clustering configuration with $N$ clusters (i.e. an $N$-element vector) and the value of each element is the product of the SINR values of all the UEs in the corresponding cluster (or partition). The immediate reward value for a state is given by the product of the sum of the rates of the users in each cluster, under the corresponding clustering configuration.
Among the different DRL methods, we investigate (i) value-based DRL methods, namely, the deep double Q-network DDQN \cite{DDQN} and State–action–reward–state–action (SARSA)~\cite{SARSA}, (ii) a policy-based DRL method, namely, the policy gradient (PG)~\cite{PG} method, and (iii) the actor-critic (AC)~\cite{AC} method. The performance, complexity, and convergence rate of these methods are then evaluated and compared. 
\\
\textbf{Value-based DRL methods (e.g. DDQN and SARSA):} 
In value-based clustering, each network partitioning configuration is assigned a certain value through a state-value function $V^{\pi}(\bm{\Gamma}_N)$, also known as the expected return function when starting at a certain state $\bm{\Gamma}_M$, where $\bm{\Gamma}_N =  (\gamma_1^{\text{P}}, \dots, \gamma_N^{\text{P}})$ and  $\gamma_n^{\text{P}} = \prod_{t = 1}^{\tau}\prod_{i_n = 1}^{\mathcal{D}_{n,j}^{\text{U}}}\gamma_{i_n}^{\{\bm{\mathcal{C}}_j\}}\left(t\right)$. The state-value function is defined as
$V^{\pi}\left(\bm{\Gamma}_N\right) = \mathbb{E}\left[r|\bm{\Gamma}_N,\pi\right],\label{Value_Function}$ 
where $r$ is the immediate reward, $\pi$ is the followed policy which can be found such that $ V^*\left(\bm{\Gamma}_N\right) = \underset{\pi}{{\text{max}}} ~V^{\pi}\left(\bm{\Gamma}_N\right),~\forall~\gamma_{n}^{\text{P}}\in \mathbb{R}, ~n = 1, \dots, N$. Given $V^*\left(\bm{\Gamma}_N\right)$, the optimal policy $\pi^*$ is found by selecting the best cell-free network partitioning configuration that maximizes $\mathbb{E}_{\bm{\Gamma}_N'  \thicksim \mathcal{T}\left(\bm{\Gamma}_N'|\bm{\Gamma}_N,\bm{\mathcal{C}}_j\right)}\left[V^*\left(\bm{\Gamma}_N'\right)\right]$, where $\mathcal{T}\left(\bm{\Gamma}_N'|\bm{\Gamma}_N,\bm{\mathcal{C}}_j\right)$ is the transition dynamics that is usually unavailable. Hence, the value function is replaced by a quality state-action-value function $Q^{\pi}\left(\bm{\Gamma}_N,\bm{\mathcal{C}}_j\right)$, which is different from $V^{\pi}$ due to the fact that a random cell-free network partitioning configuration action $\bm{\mathcal{C}}_0$ is provided and the policy $\pi$ is only counted from the succeeding state, i.e.
$ Q^{\pi}\left(\bm{\Gamma}_N,\bm{\mathcal{C}}_j\right) = \mathbb{E}\left[r|\bm{\Gamma}_N,\bm{\mathcal{C}}_j,\pi\right]$. The learning of the $Q^{\pi}$ network is performed by using the Bellman equation with the recursive form $Q^{\pi}\left(\bm{\Gamma}_N,\bm{\mathcal{C}}_j\right) = \mathbb{E}_{\bm{\Gamma}_N'}\left[r'+\zeta Q^{\pi}\left(\bm{\Gamma}_N',\pi\left(\bm{\Gamma}_N'\right)\right)\right] \label{Q_Learning}$ \cite{BellMan_Equation}.
This means that the quality function can be improved by bootstrapping (i.e. using current values of $Q^{\pi}$ to improve our estimate). This modeling is the basis of Q-learning \cite{DDQN} and SARSA \cite{SARSA} algorithms that is defined as $Q^{\pi}\left(\bm{\Gamma}_N,\bm{\mathcal{C}}_j\right) \leftarrow Q^{\pi}\left(\bm{\Gamma}_N,\bm{\mathcal{C}}_j\right) + \alpha \delta,\label{Q_Update}$ where $\alpha$ is the learning rate and $\delta = Y - Q^{\pi}\left(\bm{\Gamma}_N,\bm{\mathcal{C}}_j\right)$ is the temporal difference error with $Y$ representing a target (as in standard regression problems). Using the Q-learning cell-free network partitioning agent, the target $Y$ directly approximates $Q^*$ by setting $Y = r + \zeta \underset{j}{\text{max}}Q^{\pi}\left(\bm{\Gamma}_N',\bm{\mathcal{C}}_j\right)$ (off-policy agent), where $\zeta$ is the discount factor. On the other hand, the SARSA algorithm improves the estimate of $Q^{\pi}$ by deriving a behavioural policy from $Q^{\pi}$. This is achieved by setting $Y =  r +  \zeta Q^{\pi}\left(\bm{\Gamma}_N',\bm{\mathcal{C}}_j'\right)$ (on-policy agent). 
\\
\textbf{Policy-based DRL method for clustering:} 
In policy gradient (PG) algorithms, the modeling and optimization of a certain policy is conducted directly through a parameterized function, $\mu_{\theta}\left(\bm{\mathcal{C}}_j,\bm{\Gamma}_N\right)$. The value of the objective function (the reward) directly depends on the policy. The value of the reward function for PG-based clustering is given by
\begin{equation}
\begin{aligned}
& J(\theta) = \sum_{\bm{s}_{\text{c}}\in \mathbb{R}^N}d^{\mu}\left(\bm{\Gamma}_N\right)V^{\mu}\left(\bm{\Gamma}_N\right) = \sum_{\bm{s}_{\text{c}}\in \mathbb{R}^N}d^{\mu}\left(\bm{\Gamma}_N\right)\sum_{j\in \mathbb{Z}}\mu_{\theta}\left(\bm{\mathcal{C}}_j   |\bm{\Gamma}_N\right)Q^{\mu}\left(\bm{\Gamma}_N,\bm{\mathcal{C}}_j\right), 
\end{aligned}
\end{equation}
where $d^{\mu}\left(\bm{\Gamma}_N\right)$ is the stationary state distribution of Markov chain. Note that the gradient of $J\left(\theta\right)$ (denoted by $\nabla_{\theta}J\left(\theta\right)$) depends both on the selected actions $\bm{a}_{\text{c}}$ and the stationary distribution $d^{\mu}\left(\bm{\Gamma}_N\right)$. 
We also use a PG algorithm that  simplifies the computation of the gradient by  removing the dependence of $J\left(\theta\right)$ on $d^{\mu}\left(\bm{\Gamma}_N\right)$ as follows~\cite[Sec. 13.2]{RL_Book}:
\begin{dmath}
    \nabla_{\theta} J\left(\theta\right) = \nabla_{\theta} \sum_{\bm{s}_{\text{c}}\in \mathbb{R}^N}d^{\mu}\left(\bm{\Gamma}_N\right)\sum_{j\in \mathbb{Z}}\mu_{\theta}\left(\bm{\mathcal{C}}_j    |\bm{\Gamma}_N\right)Q^{\mu}\left(\bm{\Gamma}_N,\bm{\mathcal{C}}_j\right)\\
    \propto 
     \sum_{\bm{s}_{\text{c}}\in \mathbb{R}^N}d^{\mu}\left(\bm{\Gamma}_N\right)\sum_{j\in \mathbb{Z}}\mu_{\theta}\left(\bm{\mathcal{C}}_j    |\bm{\Gamma}_N\right)\nabla_{\theta}Q^{\mu}\left(\bm{\Gamma}_N,\bm{\mathcal{C}}_j\right).\label{PG_Gradient_1}
\end{dmath}
The general policy gradient method has a high variance. Accordingly, the following general form is used as a foundation of different PG algorithms:
\begin{dmath}
    \nabla_{\theta}J\left(\theta \right) = \mathbb{E}_{\mu_{\theta}}\left[\sum_{t = 0}^{T-1}G_t\nabla_{\theta}\log \mu_{\theta}\left(\bm{\mathcal{C}}_j|\bm{\Gamma}_N\right) \right].\label{Expected_Reward}
\end{dmath}
The PG-based DRL model for network partitioning  can be then trained through the following steps:
\begin{itemize}
    \item[i.]  Initialize the actor $\mu\left(\bm{\Gamma}_N\right)$ with random weights $\theta_{\mu}$.
    \item[ii.] For each training episode (i.e. every $T$   training steps), generate the experiences by following $\mu\left(\bm{\Gamma}_N\right)$: the actor generates the probability values for each possible cell-free partitioning, then the DRL agent randomly selects an action based on a certain probability distribution. This process continues for $T$ steps.
    \item[iii.] At each step of a certain episode, calculate the return value $G_t = \sum_{l = t}^T\zeta^{l-1}r_l$.
    \item[iv.] Find the cumulative sum of the actor network gradients during one entire learning episode as
    \begin{dmath}
        d\theta_{\mu} = \sum_{t = 1}^{T}G_t\nabla_{\theta_{\mu}}\ln \mu\left(\bm{\Gamma}_N|\theta_{\mu}\right) 
    \end{dmath}
     \item[v.] Update the actor network using: 
$\theta_{\mu} \leftarrow  \theta_{\mu} + \alpha d\theta_{\mu}$, where $\alpha$ is the learning rate.
\end{itemize}
\textbf{(Actor-critic)-based DRL method:} 
In the PG-based partitioning algorithm, the value function $G_t$ is estimated based on a preassigned policy. However, the estimation of $G_t$ for a predefined policy introduces a relatively high variance of the policy gradient  which in turn reduces the quality of cell-free network partitioning action. In order to tackle the high variance problem, a second DNN can be used that can accurately learn the value of $G_t$\cite{Discrete_AC}.
\subsection{Learn to ``Beamsteer"}
As discussed in Sec. \ref{Hybrid_Beamforming_Section}, the downlink beamforming is performed through two consecutive stages, namely, analog beamsteering and digital beamforming. 
In this section, we develop a mixed DRL-cum-convex optimization subsystem that performs the two-stage beamforming operation for each cell-free subnetwork (\textbf{Fig. \ref{DRL_1}(b)}).
In the proposed system, the non-convex analog beamsteering (problem $\textbf{\texttt{P}}_2$, which is a non-convex combinatorial optimization problem) is solved by training a DRL agent that ``interacts" with the propagation medium (i.e. DRL environment) on a time-slot basis. The convex digital beamforming problem $\textbf{\texttt{P}}_3$, which is a strictly convex optimization problem, on the other hand, is solved within the DRL environment by using conventional convex optimization methods (e.g. Newton and Broyden methods\footnote{The Lagrangian function of $\textbf{\texttt{P}}_3$ is twice differentiable w.r.t all optimization variables \cite{Broyden1965ACO}.}). This process of hybrid analog beamsteering-digital beamforming is performed independently at each subnetwork on a time-slot basis. 

\textbf{Table \ref{Table_Mixed_DRL}} shows the main design parameters of the DRL model. 
 \begin{table}[h!]
    \centering
      \caption{DRL model for hybrid beamforming in subnetwork $n$ }
\begin{tabular}{|c|c|}
\hline
\shortstack{Beamforming Environment\\ Variables} & \shortstack{Network Beamforming System Equivalence\\  \textcolor{white}{.}} \\
\hline
\hline
State $\bm{s}_{\text{b}} = \left\{{s}_{\text{b},1_n}, \dots, {s}_{\text{b},\mathcal{D}_{n,j}^{\text{U}}} \right\}$ & $\left\{\gamma_{1}^{\{\mathcal{C}_j\}}, \dots, 
\gamma_{\mathcal{D}_{n,j}^{\text{U}}}^{\{\mathcal{C}_j\}}\right\}$ (Involving Digital Beamforming)\\
\hline
Reward $r_{\text{b}}$ & $\sum_{m_n = 1}^{\mathcal{D}_{n,j}^{\text{A}}}\left(\sum_{k_n = 1}^{\mathcal{D}_{n,j}^{\text{U}}}
    ||\bm{\delta}^{\not \perp}_{k_n}\bm{\Sigma}_{k_n m_n}\bm{\mathcal{A}}_{m_n}^{\not \perp}||^2
    + \sum_{l = 1, l \neq n}^{N}\sum_{k_l = 1}^{\mathcal{D}_{l,j}^{\text{U}}}
    || \bm{\delta}^{  \perp}_{k_l}\left(t-1\right)\bm{\Sigma}_{k_lm_n}\bm{\mathcal{A}}_{m_n}^{\perp}||^2
    \right)$\\
\hline
Action $\bm{a}_{\text{b}}$ & $\left\{\bm{\delta}_{k_n}, \bm{\mathcal{A}}_{m_n}\right\}^{m_n =1, \dots, \mathcal{D}_{n,j}^{\text{A}}}_{k_n = 1, \dots, \mathcal{D}_{n,j}^{\text{U}}}$ \\
\hline
\end{tabular}
    \label{Table_Mixed_DRL}
\end{table}
Note that the operation of digital beamforming is performed as a part of the DRL environment computations that produce the observed states and reward for the analog beamsteering agent.
The proposed subsystem can be implemented by using several DRL algorithms. In this paper, we implement and benchmark two DRL algorithms with continuous action space, namely, the PG algorithm and the Soft Actor-Critic Agents (SAC) algorithm.
\\
\textbf{Policy-based beamsteering:} The PG algorithm used previously for cell-free network partitioning can be also implemented to learn the best beamsteering vectors by solving problem $\textbf{\texttt{P}}_2$. This can be achieved by optimizing over the discrete action space and then estimating the best continuous beamsteering action. Such an approximation process is relatively slow/inefficient. However, working directly with policies that emit probability distributions can increase the estimation speed of the continuous action space since sampling a well-known distribution is  easier than sampling from value functions. 
\\ 
\textbf{Soft actor-critic beamsteering:} 
On-policy actor-critic algorithms improve the stability of the network by allowing random exploration of experience from actions replay buffers \cite{RL_Book}. However, this on-policy training results in a poor sample complexity. On the other hand, off-policy algorithms have been developed to improve the sampling efficiency while maintaining robustness by developing more advanced variance reduction techniques and at the same time incorporating the off-policy samples (e.g. the DDPG family of algorithms)~\cite{lillicrap2015continuous}. However, the interaction between the off-policy DDQN value estimator and the deterministic actor setting makes
DDPG extremely difficult to stabilize and adjust the hyper-parameter settings. This issue becomes more severe as the size of the cell-free network increases. We propose to utilize the SAC algorithm to solve the beamsteering problem at each cell-free subnetwork~ \cite[Algorithm 1]{SAC_Algorithm}. The SAC algorithm enables off-policy actor-critic training with a stochastic actor. The main difference between the SAC algorithm and the GP and AC ones is that the SAC algorithm uses a general objective that maximizes entropy along with the cumulative reward \cite{Entropy_Maximization}. The addition of policy entropy encourages the actor network to explore new experiences. Accordingly, the expected reward in Eq. (\ref{Expected_Reward}) can be modified to \cite{Entropy_Maximization} 
\begin{dmath}
\nabla_{\theta_{\mu}} J\left(\theta_{\mu}\right) = \nabla_{\theta_{\mu}} \sum_{t = 0}^{T - 1} \mathbb{E}_{\left(\bm{\Gamma}_n^{\{\mathcal{C}_j\}},\bm{\mathcal{F}}_n\right)\thicksim \rho_{\theta_{\mu}}}\left[r\left(\bm{\Gamma}_n^{\{\mathcal{C}_j\}},\bm{\mathcal{F}}_n\right) + \alpha \mathcal{H}\left(\mu\left(\bm{\mathcal{F}}_n|\bm{\Gamma}_n^{\{\mathcal{C}_j\}}\right)\right) \right],
\end{dmath} 
where $\bm{\Gamma}_n^{\{\mathcal{C}_j\}} = \bm{\Gamma}_n^{\{\mathcal{C}_j\}}$, $\bm{\mathcal{F}}_n = \bm{\mathcal{F}}_n$, and $\mathcal{H}\left(.\right)$ is the entropy measure of the policy $\mu\left(\bm{\mathcal{F}}_n|\bm{\Gamma}_n^{\{\mathcal{C}_j\}}\right)$, $\alpha$ is a temperature factor that determines the relative importance of the policy entropy against the reward $r\left(\bm{\Gamma}_n^{\{\mathcal{C}_j\}},\bm{\mathcal{F}}_n\right)$, and
$\rho_{\mu}\left(\bm{\Gamma}_n^{\{\mathcal{C}_j\}}\right)$ and $\rho_{\mu}\left(\bm{\Gamma}_n^{\{\mathcal{C}_j\}},\bm{\mathcal{F}}_n\right)$ are the state and state-action of the trajectory distribution introduced by $\mu\left(\bm{\mathcal{F}}_n|\bm{\Gamma}_n^{\{\mathcal{C}_j\}}\right)$. The soft state-value function of SAC algorithm is given by \cite{SAC_Algorithm}
\begin{dmath}
V\left(\bm{\Gamma}_n^{\{\mathcal{C}_j\}}\right) =  \mathbb{E}_{\bm{a}_{\text{b}}\thicksim \bm{\mu}}\left[Q\left(\bm{\Gamma}_n^{\{\mathcal{C}_j\}},\bm{\mathcal{F}}_n\right) -  \log \mu\left(\bm{\mathcal{F}}_n|\bm{\Gamma}_n^{\{\mathcal{C}_j\}}\right)\right].\label{SVF}
\end{dmath}
Accordingly, the soft Q-value will be defined as \cite{SAC_Algorithm}
\begin{dmath}
Q\left(\bm{\Gamma}_n^{\{\mathcal{C}_j\}},\bm{\mathcal{F}}_n\right) 
= r\left(\bm{\Gamma}_n^{\{\mathcal{C}_j\}},\bm{\mathcal{F}}_n\right) + \gamma\mathbb{E}_{{\bm{\Gamma}_n^{\{\mathcal{C}_j\}}}'\thicksim \rho_{\mu}\left({\bm{\Gamma}_n^{\{\mathcal{C}_j\}}}'\right)}\left[Q\left({\bm{\Gamma}_n^{\{\mathcal{C}_j\}}}',\bm{\mathcal{F}}_n'\right) -  \log \mu\left(\bm{\mathcal{F}}_n'|{\bm{\Gamma}_n^{\{\mathcal{C}_j\}}}'\right)\right].\label{SQF}
\end{dmath}
The SAC algorithm aims to learn three functions, namely, i) a policy function with parameters $\theta$ and $\pi_{\theta}$, ii) a soft Q-value function approximated (parameterized) by $w$ and $Q_w$, and iii) a soft state value function parameterized by $\psi$ and $V_{\psi}$. The soft state value is trained to minimize the mean square error with gradient function given as follows \cite{SAC_Algorithm}:
\begin{dmath}
\nabla_{\psi} J_V\left(\psi\right) = \nabla_{\psi} \mathbb{E}_{\bm{\Gamma}_n^{\{\mathcal{C}_j\}}\thicksim \mathcal{R}}\left[\frac{1}{2}\left(V_{\psi}\left(\bm{\Gamma}_n^{\{\mathcal{C}_j\}}\right) 
- 
\mathbb{E} \left[Q_{w}\left(\bm{\Gamma}_n^{\{\mathcal{C}_j\}},\bm{\mathcal{F}}_n\right) 
- \log \pi_{\theta}\left(\bm{\mathcal{F}}_n|\bm{\Gamma}_n^{\{\mathcal{C}_j\}}\right)\right]\right)^2\right]\\
\approx \nabla_{\psi}V_{\psi}\left(\bm{\Gamma}_n^{\{\mathcal{C}_j\}}\right) \left(V_{\psi}\left(\bm{\Gamma}_n^{\{\mathcal{C}_j\}}\right) - Q_w\left(\bm{\Gamma}_n^{\{\mathcal{C}_j\}},\bm{\mathcal{F}}_n\right) + \log \pi_{\theta}\left(\bm{\mathcal{F}}_n|\bm{\Gamma}_n^{\{\mathcal{C}_j\}}\right)\right),
\label{SVF_Gradient}
\end{dmath}
where $\mathcal{R}$ is the distribution of previously sampled actions and states (in the replay buffer). Furthermore, the soft Q function is trained to minimize the soft Bellman residual with gradient function given as 
\begin{dmath}
\nabla_{w}J_Q\left(w\right) = \nabla_{w} \mathbb{E}_{\left(\bm{\Gamma}_n^{\{\mathcal{C}_j\}},\bm{\mathcal{F}}_n\right)\thicksim \mathcal{R}}\left[\frac{1}{2}\left(Q_{w}\left(\bm{\Gamma}_n^{\{\mathcal{C}_j\}},\bm{\mathcal{F}}_n\right)
- \left(r\left(\bm{\Gamma}_n^{\{\mathcal{C}_j\}},\bm{\mathcal{F}}_n\right) 
\\
+ \zeta \mathbb{E}_{{\bm{\Gamma}_n^{\{\mathcal{C}_j\}}}'\thicksim \rho_{\pi}\left(\bm{\Gamma}_n^{\{\mathcal{C}_j\}}\right)}\left[ V_{\bar{\psi}}\left({\bm{\Gamma}_n^{\{\mathcal{C}_j\}}}'\right)\right] \right) \right)^2 \right]\\ 
\approx \nabla_{w}  Q\left(\bm{\Gamma}_n^{\{\mathcal{C}_j\}},\bm{\mathcal{F}}_n\right)
\left(Q_w\left(\bm{\mathcal{F}}_n,\bm{\Gamma}_n^{\{\mathcal{C}_j\}}\right)
- r\left(\bm{\Gamma}_n^{\{\mathcal{C}_j\}},\bm{\mathcal{F}}_n\right) - \zeta V_{\bar{\psi}}\left({\bm{\Gamma}_n^{\{\mathcal{C}_j\}}}'\right)\right), 
\label{SQF_Gradient}
\end{dmath}
where $\bar{\psi}$ is an exponentially moving average target function. The desired policy is then trained using the information projection that is defined in terms of Kullback-Leibler (KL)-divergence \cite{KL_Divergence}. Accordingly, the policy is updated according to
\begin{dmath}
\pi_{\text{new}} = \argmax_{\pi'\in \Pi}D_{\text{KL}}\left(\pi'\left(.|\bm{\Gamma}_n^{\{\mathcal{C}_j\}}\right)||\frac{\exp\left\{Q^{\pi_{\text{old}}}\left(\bm{\Gamma}_n^{\{\mathcal{C}_j\}},.\right)\right\}}{Z^{\pi^{\text{old}}}\left(\bm{\Gamma}_n^{\{\mathcal{C}_j\}}\right)}\right),
\end{dmath}
where $\Pi$ denotes a set of potential policies that $\pi$ must restricted to. $Z^{\pi^{\text{old}}}\left(\bm{\Gamma}_n^{\{\mathcal{C}_j\}}\right)$ is a partitioning function that is used for normalizing the distribution. The objective update function of the policy $\pi_{\theta}$ is~\cite{SAC_Algorithm}
\begin{dmath}
\nabla_{\theta} J_{\pi}\left(\theta\right) = D_{\text{KL}}\left(\pi_{\theta}\left(.|\bm{\Gamma}_n^{\{\mathcal{C}_j\}}\right)||\exp\left\{Q_w\left(\bm{\Gamma}_n^{\{\mathcal{C}_j\}},\bm{\mathcal{F}}_n\right)   
-\log Z_w\left(\bm{\Gamma}_n^{\{\mathcal{C}_j\}}\right)\right\}\right)\\
= \mathbb{E}_{\bm{\mathcal{F}}_n\thicksim \pi}\left[-\log \left(\frac{ 
    \exp\left\{Q_w\left(\bm{\Gamma}_n^{\{\mathcal{C}_j\}},\bm{\mathcal{F}}_n\right)
    - \log Z_w\left(\bm{\Gamma}_n^{\{\mathcal{C}_j\}}\right)\right\}}{\pi_{\theta}\left(\bm{\mathcal{F}}_n|\bm{\Gamma}_n^{\{\mathcal{C}_j\}}\right)}\right)\right]\\
= \mathbb{E}_{\bm{\mathcal{F}}_n\thicksim \pi}\left[ \log \pi_{\theta}\left(\bm{\mathcal{F}}_n|\bm{s}_{\text{b}}\right) 
-Q_w\left(\bm{\Gamma}_n^{\{\mathcal{C}_j\}},\bm{\mathcal{F}}_n\right) + \log Z_w\left(\bm{\Gamma}_n^{\{\mathcal{C}_j\}}\right)\right].
\end{dmath}
We use the SAC algorithm developed in \cite[Algorithm 1]{SAC_Algorithm}.
\\

 \textbf{Algorithm \ref{Systm_Model_Algorithm}} shows the sequence of processes performed during network operation, where $E_{\text{c}}$ and $E_{\text{b}}$ are the number of episodes for the DRL models used for clustering and beamsteering, respectively, $T_{\text{c}}$ and $T_{\text{b}}$ are the number of training steps in each episode for the DRL models used for clustering and beamsteering, respectively. 
\begin{algorithm}
\caption{Training for network partitioning and beamforming}
\label{algo:ddpg}
\begin{algorithmic}[1]
\State Input: $\bm{H_{k_n m_n}}$, $M$, $K$, $a$, $u$, and $N$.
\State Initialize target networks of main DRL system and subsystems in \textbf{Figs. \ref{DRL_1}(a)} and \textbf{(b)}\footnotemark.
\State Initialize replay buffers of the main DRL system and the subsystems in \textbf{Figs. \ref{DRL_1}(a)} and \textbf{(b)}.   
\State Initialize $\bm{\mathcal{C}}_j$, $\bm{\mathcal{A}}_{m_n}$, $\bm{\delta}_{k_n}$ and $\bm{w}_{k_n m_n}$, $\forall~n, m_n$, and $k_n$.
\For{$\text{Episodes}_{\text{Clustering}} = 1\; \mbox{to}\; E_{\text{c}}$}
\For{$t_{\text{c}} = 1\; \mbox{to}\; T_{\text{c}}$} \Comment{\underline{Training steps for network partitioning}}
\State Update weights of target networks using Algorithms in \textbf{Sec. V.B} 
 
\ParFor{$n = 1\; \mbox{to}\; N$}\Comment{\underline{Simultaneous beamforming in all subnetworks}}
    
\For{$\text{Episodes}_{\text{Beamsteering}} = 1\; \mbox{to}\; E_{\text{b}}$}
\For{$t_b = 1\; \mbox{to}\; T_b $}

\State Update weights of target networks using the Algorithms in \textbf{Sec. V.C}.
\State Compute $\bm{\mathcal{H}}_{k_n m_n},~\forall~k_n = 1, \dots, \mathcal{D}_{n,j}^{\text{U}}$ and $m_n = 1, \dots, \mathcal{D}_{n,j}^{\text{A}}$.
\State Solve problem $\textbf{\texttt{P}}_3$ using a convex optimizer.
\State Update $\bm{\mathcal{A}}_{m_n}$, $\bm{\delta}_{k_n}$, and $\bm{w}_{k_n m_n}$, $\forall~n, m_n$, and $k_n$.
\EndFor
\EndFor
    
\EndParFor
\State Update $\bm{\mathcal{C}}_j$.


\EndFor
\EndFor
\end{algorithmic}\label{Systm_Model_Algorithm}
\end{algorithm}
\footnotetext{The type of network will depend on the utilized clustering and beamsteering techniques (\textbf{Sec. V.B and Sec. V.C}).}.   


\section{Complexity Analysis and Signaling Overhead}\label{Complexity}
To solve problem $\textbf{\texttt{P}}_1$ in (\ref{Opt_Prob_1}), the following subproblems will need to be solved: the combinatorial problem related to selecting the best network partitioning configuration, the non-convex problem related to finding the best beamsteering matrices (problem $\textbf{\texttt{P}}_2$ in (\ref{Opt_Prob_2})), and a convex optimization problem related to finding the optimal digital beamforming at each eAP. 

In \textbf{Sec. \ref{Learn_to_Cluster}}, we discussed how the complexity of finding best network partitions grows exponentially with increasing values of $M$, $K$, and $N$, as can be seen from (\ref{eq:bell-number}). Furthermore, solving $\textbf{\texttt{P}}_2$ in (\ref{Opt_Prob_2}) through an exhaustive search with a step size $\Delta$ will have a complexity of order $O\left(\prod_{n = 1}^N\left(\frac{1}{\Delta}\right)^{(a\times \mathcal{D}_{n,j}^{\text{A}})\times (u\times \mathcal{D}_{n,j}^{\text{U}})}\right)$. Since the problem $\textbf{\texttt{P}}_3$ in (\ref{Opt_Prob_3}) is strictly convex, the solution for this problem has a computational complexity of $O\left(\left(\sum_{n = 1}^N\mathcal{D}_{n,j}^{\text{U}}\times \mathcal{D}_{n,j}^{\text{A}} \right)^3\right)$.

To evaluate the time-complexity of a deep neural network used in a DRL model, the conventional measure is the \textit{floating-point operations per second} (FLOPs). For any fully connected layer $\mathbb{L}_i$ of input size $I_i$ and output size $O_i$, the number of FLOPs is given by $\text{FLOPS}(\mathbb{L}_i) = 2 \,I_i\,O_i$. 
The policy network has two hidden layers of size 256 and 128.  Thus, for the DRL models, the total number of FLOPS during the inference is
\begin{equation}
\begin{aligned}
\text{FLOPs}_{\text{DRL}}&=\sum_{i=1}^{3}\, \text{FLOPs}(\mathbb{L}_i) 
&= 2 \cdot \Big( 256 \cdot\mathcal{|S|} + 128 \cdot \bm{|A|} + 32768 \Big),
\end{aligned}
\end{equation}
where $\mathcal{|S|}$ and $|\bm{A}|$ are the dimensions of the state space and action space, respectively.
\textbf{Table \ref{Comparision}} compares the  FLOPS for inference as well as the convergence rate for the DRL algorithms used in this paper.
\begin{table}[h!]\scriptsize
\centering
\caption{Complexity of different DRL models for clustering}
\begin{tabular}{|c||c|c|}
\hline 
\textbf{DRL Agent} & \textbf{Inference FLOPS} & \textbf{Convergence} \\
\hline
\hline
\shortstack{Conventional solution\\ \textcolor{white}{.}} 
 & \shortstack{$\Theta \left(M,K,N\right)$\\ \textcolor{white}{ }}
& \shortstack{Linear\\ convergence} \\
\hline
SARSA 
& $32768 + 256\cdot{K} + 128$  
& Slow \\
\hline
DDQN 
& $32768 + 256\cdot{K} + 128$  
& Geometric \\
\hline 
PG 
& $32768 + 256\cdot{K} + 128$  
& Sub-linear \\
\hline
Actor-Critic 
& $2 \left(32768 + 256\cdot{K} + 128\right)$  
& Fast \\
\hline
\end{tabular}
\label{Comparision}
\end{table}
Note that for network clustering, the dimensionality of the action space is $|\bm{A}| = 1$. Similarly, the complexity and convergence properties for the considered beamsteering agents are summarized in \textbf{Table \ref{Comparision_2}}\footnote{The ellipsoid method requires a total of $O\left(\left[\left((\mathcal{D}_{n,j}^{\text{A}}\times \mathcal{D}_{n,j}^{\text{U}}\right)\times \left(a\times u\right)\right]^4q \right)$ operations, where $q$ is the length of binary coding of the input.}. 
\begin{table}[h!]\scriptsize
\centering
\caption{Complexity of DRL models for beamsteering in the $n$-th subnetwork}
\begin{tabular}{|c||c|c|}
\hline 
\textbf{DRL Agent} & \textbf{Inference FLOPS} & \textbf{Convergence} \\
\hline
\hline
\shortstack{Conventional solution\\ \textcolor{white}{.}} 
 & \shortstack{$O\left(\left[\left(\mathcal{D}_{n,j}^{\text{A}}\times \mathcal{D}_{n,j}^{\text{U}}\right)\times \left(a\times u\right)\right]^2q \right)$\\ \textcolor{white}{ }}
& \shortstack{Linear\\ convergence} \\
\hline
PG 
& $32768 + 256\cdot{K} + 128\cdot{\mathcal{M}_n}$  
& Sub-linear \\
\hline
DDPG 
& $32768 + 256\cdot{K} + 128\cdot{\mathcal{M}_n}$  
& Unknown \\
\hline 
SAC
& $32768 + 256\cdot{K} + 128\cdot{\mathcal{M}_n}$  
& Unknown \\
\hline
\end{tabular}
\label{Comparision_2}
\end{table}
$\mathcal{M}_n$ in \textbf{Table \ref{Comparision_2}} represents the dimensionality of the analog beamsteering problem and is given by $
\mathcal{M}_n = \left(\mathcal{D}_{n,j}^{\text{A}}\times \mathcal{D}_{n,j}^{\text{U}}\right)\times \left(a\times u\right)$.


{In terms of signaling overhead, in the proposed methods, the NCC first will have to collect the estimated CSI matrices from distributed eAPs and send full CSI to the ECP of each cell-free subnetwork. Next, the NCC will collect the performance metric (e.g. sum-rate) from the ECP of each subnetwork and use it to decide on the new network partitioning configuration.}


\section{Numerical Results} 
 
\subsection{Parameters and Assumptions}
\textbf{Table \ref{Simulation_Parameters}} presents the values of different parameters used in generating the simulation results. 
\begin{table}[h!]
    \centering
      \caption{Simulation parameters}
\begin{tabular}{|c|c|}
\hline 
Parameter & Value  \\
\hline
\hline
AWGN PSD at UE &
$-169$ dBm/Hz  
\\
\hline
Path-loss exponent & 2 (outdoor)
\\
\hline
mmWave carrier frequency, $\frac{3\times 10^8}{\lambda}$ & $24$ GHz (unless specified otherwise)
\\
\hline
mmWave paths, $\mathcal{L}$ & $3$ (unless  specified otherwise )
\\
\hline 
SIC sensitivity, $P_s$  & $1$ dBm
\\
\hline
$\#$ of training episodes & $\{2000, 4000\}$
\\
\hline
$\#$ of training steps/episode & $200$
\\
\hline
Discount factor, $\zeta$ & $0.01$
\\
\hline
Learning rate, $\alpha$ & $0.001$
\\
\hline
\end{tabular}
    \label{Simulation_Parameters}
\end{table}
All the results for the conventional methods are produced using Mont-Carlo simulations each with $10^6$ runs. Additionally, we assume that all channel small-scale fading gains $h_{k_n m_n}$ are drawn from a set of i.i.d random variables. We assume that all APs and UEs are uniformly distributed over a disc of radius 18 m (corresponding to a network total coverage area of $1 \text{km}^2$).
 
\subsection{Hybrid Beamforming Scheme}
\begin{figure}[htb]
		\centering
		\includegraphics[scale = 0.82]{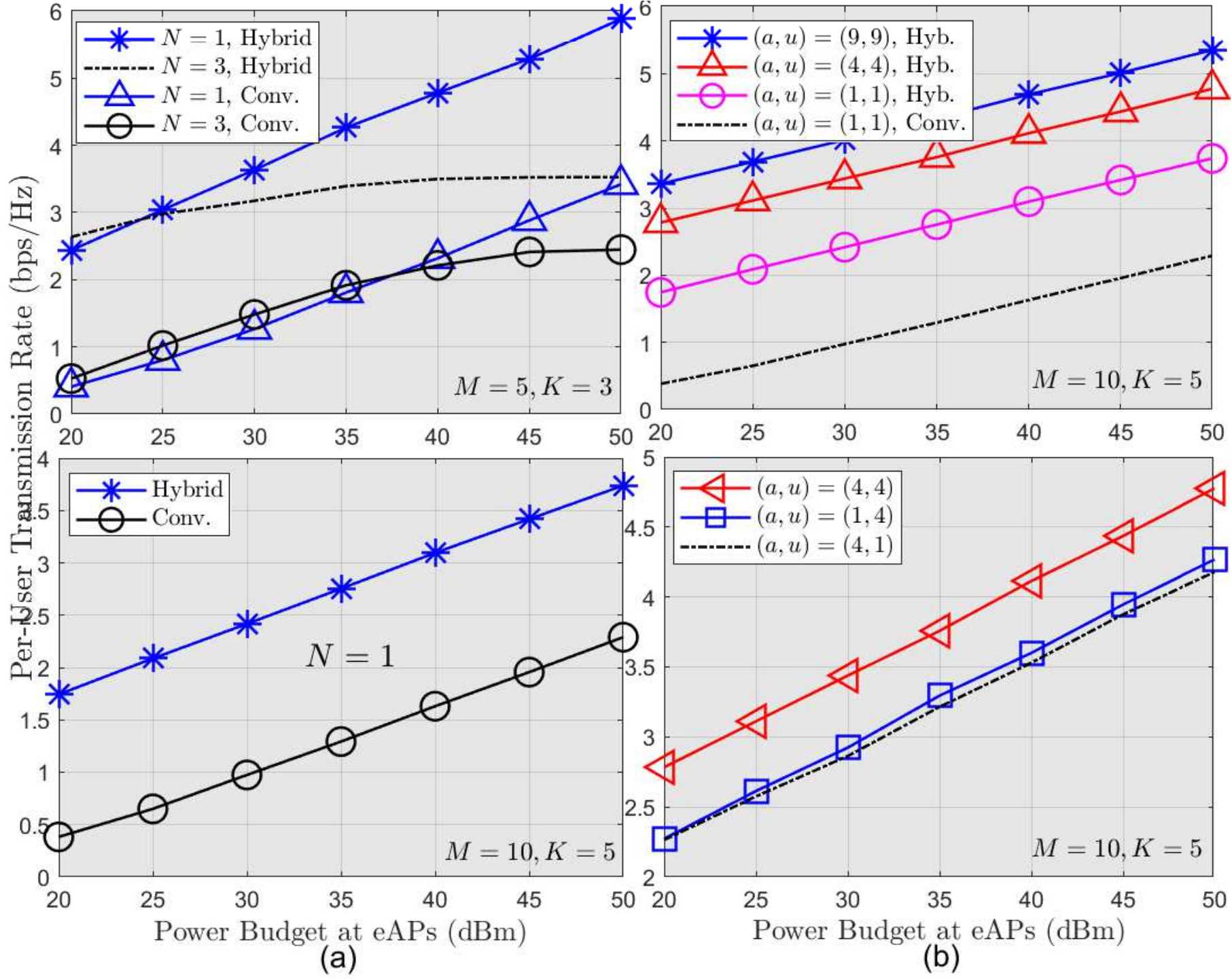}
		\caption{Hybrid vs. conventional beamforming techniques.}\label{Hybrid_Vs_Conventional}
	\end{figure}

We start this section by evaluating the performance of the proposed network architecture under the designed hybrid beamforming system (\textbf{Fig. \ref{Hybrid_Vs_Conventional}(a)}).
It can be noticed from this figure that the designed hybrid analog beamsteering-digital beamforming scheme significantly outperforms that of conventional (all digital) beamforming scheme. For example, a gain of $2.3$ bps/Hz and $1.5$ bps/Hz are achieved at $35$ dBm with $N = 1$ and $N = 2$, respectively (upper graph in \textbf{Fig. \ref{Hybrid_Vs_Conventional}(a)}). Interestingly, even without network partitioning (i.e. for $N = 1$), our proposed hybrid beamforming scheme shows significant increase in performance compared to its conventional beamforming counterpart. The reason is that our designed objective function for beamsteering (see \textbf{Problem $\texttt{P}_2$} in (\ref{Opt_Prob_2})) aims to enhance beams of desired UEs and ``\textit{zero-null}" beams to undesired UEs at the same time. Accordingly, with $N = 1$, beamsteering will focus on optimally directing the antenna main lobes of APs and UEs  to each other. However, this performance gain is observed to decrease as the network scales up (see lower graph of \textbf{Fig. \ref{Hybrid_Vs_Conventional}(a)}).   
In order to study the effect of multiple antennas on the per-UE rate performance, \textbf{Fig. \ref{Hybrid_Vs_Conventional}(b)} shows per-UE transmission rate versus different MIMO layouts.
It can be noticed that a significant increase in per-UE rate performance can be achieved by increasing the number of antennas at the UEs and eAPs (upper graph of \textbf{Fig. \ref{Hybrid_Vs_Conventional}(b)}). This rate enhancement, however, decreases as the values of $a$ and/or $b$ increase (due to increased interference levels). 
Furthermore, the system performance enhances better as the antenna order at the UEs increases more than that at the eAPs (lower graph of \textbf{Fig. \ref{Hybrid_Vs_Conventional}(b)}). 
\subsection{Evaluation and Benchmarking of Hierarchical DRL Models}
The performances of the DRL models for network partitioning and analog beamsteering are investigated separately. This is done by first training the different DRL clustering agents while using conventional methods for hybrid analog beamsteering-digital beamforming operations. On the other hand, the DRL-based beamforming methods are evaluated while clustering is performed through the trained DRL agents in the inference mode. This separate evaluation enables us to extract more insights on the performances of that DRL models under discrete and continuous action spaces. 

\begin{figure}[htb]
		\centering
		\includegraphics[scale = 0.70]{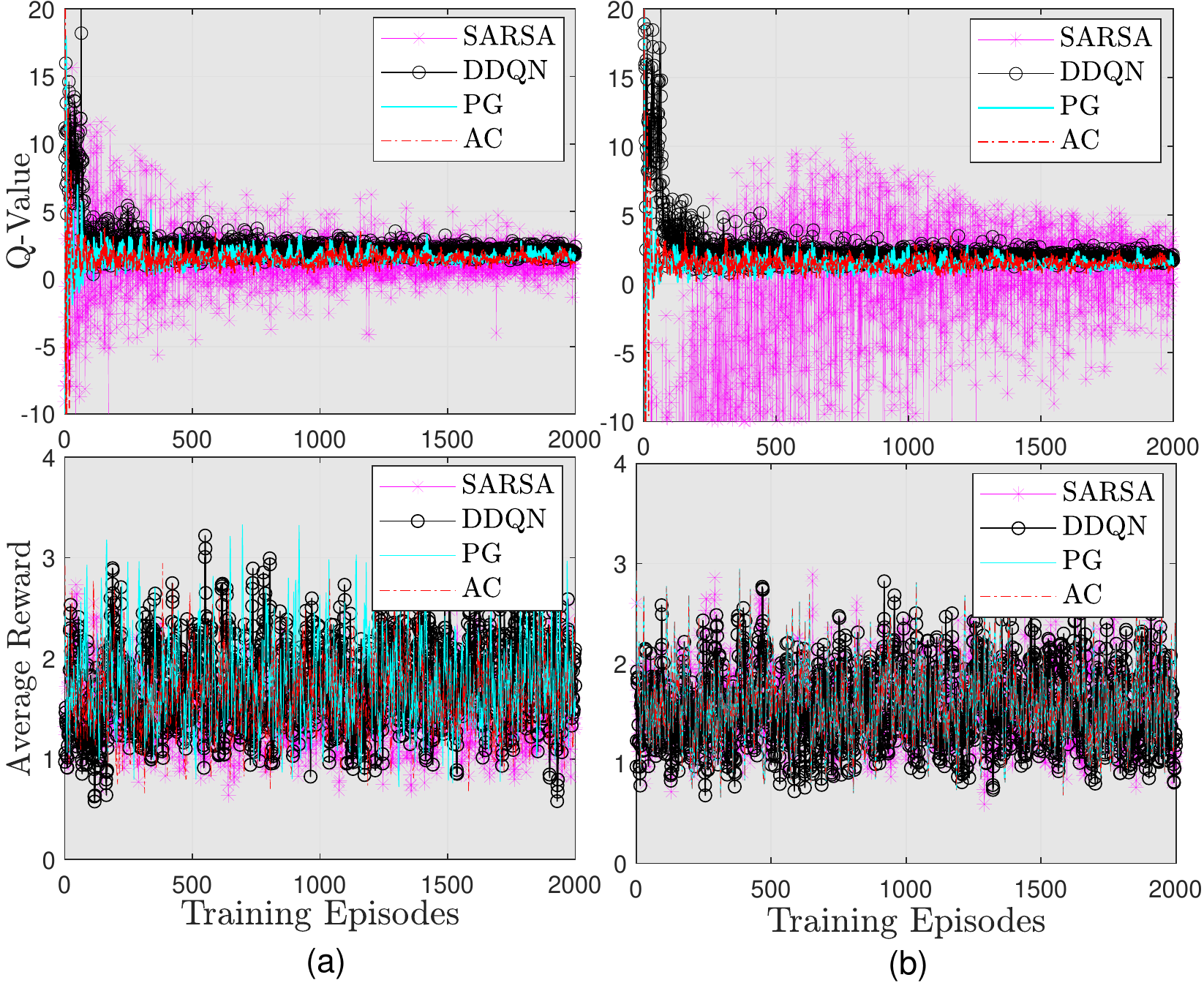}
		\caption{Performance of clustering agents: (a) fixed CSI, and (b) varying CSI.}\label{Clustering_Algorithms_Comparision}
	\end{figure}

We start by evaluating the performances of the DRL-based clustering algorithms for the proposed self-partitioning cell-free network architecture (\textbf{Fig.  \ref{Clustering_Algorithms_Comparision}(a)}). We use two training modes for each of the studied DRL algorithms. The first training mode considers a fixed CSI (i.e. constant $\bm{H}$ matrix), while in the second mode, we use different CSI realizations at every training step of each episode.
In \textbf{Fig. \ref{Clustering_Algorithms_Comparision}(a)}, we train four DRL agents using PG, DDQN, SARSA, and AC algorithms for network partitioning for a single CSI realization. 
As can be observed, the PG algorithm provides the best clustering performance in terms of stability and convergence, while the DDQN algorithm comes second, and the SARSA algorithm comes last with significantly high covariance in the Q-values per episode and slower convergence rate (upper graph in \textbf{Fig. \ref{Clustering_Algorithms_Comparision}(a)}). In terms of per-UE rate performance, even though all of the three algorithms show relatively similar performance levels, however, with a closer look, one can find out that the PG algorithm provides the highest per-UE transmission rate.

\begin{figure}[htb]
		\centering 
		\includegraphics[scale = 0.70]{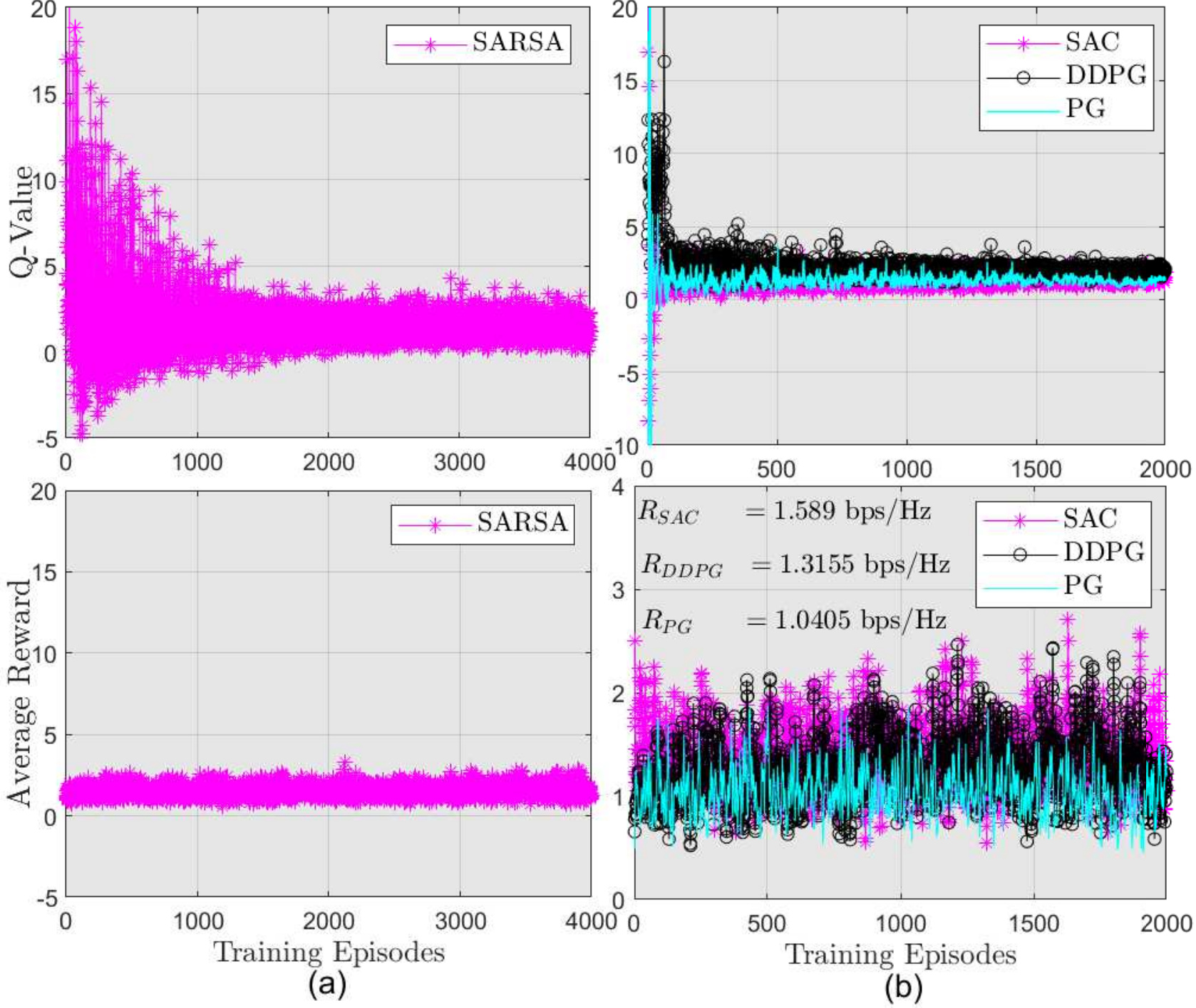}
		\caption{Performance of different clustering agents with different number of training episodes.}\label{SARSA_Pluse_Beamsteering}
\end{figure}

\textbf{Fig. \ref{Clustering_Algorithms_Comparision}(b)} evaluates the effect of training DRL agents during the real-time operation of the cell-free network. Specifically, we assume that a training step is performed during one time slot. This means that, state transitions of the DRL model will result from both current action $\bm{a}_c$ and the instantaneous CSI $\bm{H}$. 
It can be observed from \textbf{Fig. \ref{Clustering_Algorithms_Comparision}(b)} that changing $\bm{H}$ during training of the clustering agents has a negative impact on both the convergence rate and the per-UE rate performance. This can be observed clearly by the significant increase in the variance of the Q-values in the upper graph of \textbf{Fig. \ref{Clustering_Algorithms_Comparision}(b)}. It can also be observed that the convergence of the SARSA-based clustering is the worst. For SARSA, to tackle the weak stability issue, we double the number of training episodes from 2000 to 4000 episodes and retrain the SARSA agent under varying CSI conditions (\textbf{Fig. \ref{SARSA_Pluse_Beamsteering}(a)}).  
As can be observed from this figure, increasing the number of episodes improves the stability of the SARSA algorithm significantly. However, with more training episodes the per-UE rate performance does not improve (with average reward of around $ 1.5408$ bps/Hz). More numerical results on the performances of the DRL-based clustering schemes are given in \textbf{Table \ref{DRL_Clustering_Agents_Table}}. 
\begin{table}[h!]
    \centering
      \caption{Numerical results on the performances of different clustering schemes.}
\begin{tabular}{|c|c|c|c|c|}
\hline 
\multicolumn{5}{|l|}{\hspace{12mm}$(M,K) = (5,3)$ and $(a,u) = (1,1)$ with optimal performance: $2.03786$ pbs/Hz}\\
\hline 
\shortstack{ Agent \\ \textcolor{white}{ } } 
& \shortstack{Average reward \\ \textcolor{white}{ }} 
& \shortstack{Inference mode \\ Fixed $\bm{H}$} 
&  \shortstack{Inference mode \\ Variable $\bm{H}$} 
& \shortstack{Training duration \\ (2000 episodes)} 
\\
\hline
\hline
\shortstack{PG trained by\\ fixed $\bm{H}$} 
& \shortstack{$1.7607$ bps/Hz\\ \textcolor{white}{ }} 
& \shortstack{$ 1.5626$ bps/Hz\\ \textcolor{white}{ }} 
& \shortstack{$1.4841$ bps/Hz\\ \textcolor{white}{ }} 
& \shortstack{$15.4072$ Mins\\ \textcolor{white}{ }}  \\
\hline
\shortstack{PG trained by\\ varying $\bm{H}$} 
& \shortstack{$1.6090$ bps/Hz\\ \textcolor{white}{ }} 
& \shortstack{$1.5592$ bps/Hz\\ \textcolor{white}{ }} 
& \shortstack{$1.5332$ bps/Hz\\ \textcolor{white}{ }} 
& \shortstack{$15.0098$ Mins\\ \textcolor{white}{ }}   \\
\hline
\shortstack{DDQN trained by\\ fixed $\bm{H}$}  & \shortstack{$1.7308$ bps/Hz\\ \textcolor{white}{ }}  
& \shortstack{$1.5802$ bps/Hz\\ \textcolor{white}{ }}
& \shortstack{$1.5355$ bps/Hz\\ \textcolor{white}{ }} 
& \shortstack{$16.5467$ Mins\\ \textcolor{white}{ }} 
\\
\hline
\shortstack{DDQN trained by\\ varying $\bm{H}$} 
& \shortstack{$1.5579$ bps/Hz \\ \textcolor{white}{ }} 
& \shortstack{$ 1.5321$ bps/Hz \\ \textcolor{white}{ }}   
& \shortstack{$1.4654$ bps/Hz \\ \textcolor{white}{ }}  
& \shortstack{$14.0558$ Mins \\ \textcolor{white}{ }}\\
\hline
\shortstack{SARSA trained by\\ fixed $\bm{H}$}
& \shortstack{$1.4733$ bps/Hz \\ \textcolor{white}{ }}   
&  \shortstack{$1.4376$ bps/Hz \\ \textcolor{white}{ }}   
& \shortstack{$1.5153$ bps/Hz \\ \textcolor{white}{ }}  
&  \shortstack{ $18.2317$ Mins \\ \textcolor{white}{ }}  \\
\hline
\shortstack{SARSA trained by\\ varying $\bm{H}$}  
& \shortstack{$1.5862$ bps/Hz\\ \textcolor{white}{ }}  
& \shortstack{$1.5047$ bps/Hz\\ \textcolor{white}{ }}
& \shortstack{$1.5072$ bps/Hz\\ \textcolor{white}{ }}
&   $16.3313$ Mins. \\
\hline
\shortstack{AC trained by\\ fixed $\bm{H}$}
& \shortstack{$1.4186$ bps/Hz \\ \textcolor{white}{ }}   
&  \shortstack{$1.5876$ bps/Hz \\ \textcolor{white}{ }}   
& \shortstack{$1.4991$ bps/Hz \\ \textcolor{white}{ }}  
&  \shortstack{ $15.3618$ Mins \\ \textcolor{white}{ }}  \\
\hline
\shortstack{AC trained by\\ varying $\bm{H}$}  
& \shortstack{$1.6084$ bps/Hz\\ \textcolor{white}{ }}  
& \shortstack{$1.5664$ bps/Hz\\ \textcolor{white}{ }}
& \shortstack{$1.4919$ bps/Hz\\ \textcolor{white}{ }}
& \shortstack{$15.5362$ Mins\\ \textcolor{white}{ }}    \\
\hline
\end{tabular}
    \label{DRL_Clustering_Agents_Table}
\end{table}
It can be observed that the off-policy algorithm (i.e. DDQN) gives the worst performance under varying CSI conditions. This is due to the fact that the DDQN agent selects the action related to the highest Q-value in a deterministic fashion, without any exploration. This action-selection strategy will prevent the network from learning/sensing the stochastic variations of the states (or alternatively the CSI matrix $\bm{H}$).  On the other hand, all of the on-policy based algorithms (PG, SARSA, and AC) show good performance under varying CSI, and we observe a noticeable enhancement on per-UE rate performance due to the AC algorithm. The reason is that the AC algorithm allows the DRL agent to learn the stochastic properties of the state (or alternatively, the CSI matrix $\bm{H}$).  
Finally, \textbf{Fig. \ref{SARSA_Pluse_Beamsteering}(b)} evaluates and compares the performance of several DRL-based beamsteering methods. For these simulations, we use the inference mode of the DDQN algorithm to solve the network partitioning  problem.
It can be noticed that the AC algorithm shows the best per-UE rate performance compared to the DDPG and PG algorithms. Furthermore, when both network clustering and analog beamsteering are implemented through the DRL agents, the per-UE rate performance of the SAC algorithm drops to around $70\%$ of the optimal performance, and for the PG algorithm, it is $51\%$ of the optimal performance.  
\section{Conclusion}
A novel self-partitioning MIMO cell-free network architecture has been proposed in which a cell-free network is partitioned into a set of independent cell-free subnetworks. To efficiently solve the problem of network partitioning for a large-scale network, we have proposed, evaluated, and benchmarked several state-of-the-art DRL methods with discrete action space. Furthermore, to reduce the interference between adjacent cell-free subnetworks, we have designed a novel downlink hybrid analog beamsteering-digital beamforming scheme. 
We have also evaluated several state-of-the-art DRL methods with continuous action space to solve the combinatorial problem of analog beamsteering while the digital beamforming problem has been solved as a strictly convex optimization problem. Results have showed a significant rate enhancement and complexity reduction due to the proposed hybrid beamforming scheme compared to its conventional all-digital counterpart. It has been observed that online training of different DRL agents is only slightly affected by changing the CSI in the network. However, changing the CSI can significantly affect the variance and convergence rate of the DRL algorithms such as the SARSA algorithm. Furthermore, it has also been noticed that all DRL methods for network clustering and beamsteering  have almost the same per-UE rate performance with a slight superiority of the PG and AC algorithms when used for network clustering and analog beamsteering, respectively. A potential extension of this work is to enable distributed beamforming at each cell-free subnetwork. This may be achieved by utilizing a multiple agent algorithm with continuous action space. Another extension of this work is to solve the problem of pilot assignment using distributed multiple agent DRL modeling.


 
 
\bibliographystyle{IEEEtran}
\bibliography{IEEEabrv,yasser}

\end{document}